\documentclass[twocolumn,aps,showpacs,floatfix,prc]{revtex4}
\usepackage[dvips]{epsfig}
\begin{document}

\title{From nuclear reactions to compact stars: a unified approach}

\author{D.N. Basu$^1$, Partha Roy Chowdhury$^2$, Abhishek Mishra$^3$ }
\affiliation{ $^1$Variable  Energy  Cyclotron  Centre, 1/AF Bidhan Nagar, Kolkata 700 064, India }
\affiliation{ $^2$Dept.of Physics, Govt.Degree College, Kamalpur, Dhalai, Tripura 799 285, India }
\affiliation{ $^3$Reactor Physics and Nuclear Engineering Section, RRS Division, Bhabha Atomic Research Centre, Mumbai 400 085, India }

\email[E-mail 1: ]{dnb@vecc.gov.in}
\email[E-mail 2: ]{royc.partha@gmail.com}
\email[E-mail 3: ]{mishraabhishek7@gmail.com}

\date{\today }

\begin{abstract}

    An equation of state (EoS) for symmetric nuclear matter is constructed using the density dependent M3Y effective interaction and extended for isospin asymmetric nuclear matter. Theoretically obtained values of symmetric nuclear matter incompressibility, isobaric incompressibility, symmetry energy and its slope agree well with experimentally extracted values. Folded microscopic potentials using this effective interaction, whose density dependence is determined from nuclear matter calculations, provide excellent descriptions for proton, alpha and cluster radioactivities, elastic and inelastic scattering. The nuclear deformation parameters extracted from inelastic scattering of protons agree well with other available results. The high density behavior of symmetric and asymmetric nuclear matter satisfies the constraints from the observed flow data of heavy-ion collisions. The neutron star properties studied using $\beta$-equilibrated neutron star matter obtained from this effective interaction for pure hadronic model agree with the recent observations of the massive compact stars such as PSR J1614-2230, but if a phase transition to quark matter is considered such agreement is no longer possible. 
    
\end{abstract}


\maketitle

\section{Introduction}

    The measurements of nuclear masses, densities and collective excitations have allowed to resolve some of the basic features of the equation of state (EoS) of nuclear matter. However, the symmetry properties of the EoS due to differing neutron and proton numbers remain more elusive to date and study of the isospin dependent properties of asymmetric nuclear matter and the density dependence of the nuclear symmetry energy (NSE) have become the prime objective \cite{Li08,St05,Ba05,Pi05,Ch05,SL05}. The new radioactive ion beam facilities provide the possibility of exploring the properties of nuclear matter and nuclei under the extreme condition of large isospin asymmetry. Consequently, the ultimate goal of such study is to extract information on the isospin dependence of in-medium nuclear effective interactions as well as the EoS of isospin asymmetric nuclear matter, particularly its isospin-dependent term or the density dependence of the NSE. This knowledge, especially the latter, is important for understanding not only the structure of radioactive nuclei, the reaction dynamics induced by rare isotopes and the liquid-gas phase transition in asymmetric nuclear matter, but also many critical issues in astrophysics \cite{Li08,St05,Ba05,Da02}. 

    In this work, based on the theoretical description of nuclear matter using the density dependent M3Y-Reid-Elliott effective interaction \cite{Be77,Sa79} (DDM3Y), we carry out a systematic study of the symmetric nuclear matter (SNM) and isospin-dependent bulk properties of asymmetric nuclear matter. In particular, we study the density dependence of the NSE and extract the slope $L$ and the curvature $K_{sym}$ parameters of the NSE and the isospin dependent part $K_\tau$ of the isobaric incompressibility. We compare the results with the constraints recently extracted from analyses of the isospin diffusion data from heavy-ion collisions based on the isospin and momentum-dependent IBUU04 transport model with in-medium nucleon-nucleon (NN) cross sections \cite{Ch05,Li05}, isoscaling analyses of isotope ratios in intermediate energy heavy-ion collisions \cite{SY07} and measured isotopic dependence of the giant monopole resonances (GMR) in even-A Sn isotopes \cite{Li07} and from the neutron skin thickness of nuclei \cite{Ce09}. 

    The lifetimes of radioactive decays are calculated theoretically within the improved WKB approximation \cite{ke35} using microscopic proton, $\alpha$ and nucleus-nucleus interaction potentials. These nuclear potentials have been obtained by folding the densities of the emitted and the daughter nuclei with the M3Y effective interaction, whose density dependence is determined from nuclear matter calculations. These calculations provide reasonable estimates of half-lives for the observed proton \cite{BCS08}, $\alpha$ \cite{Ba03,CSB06,prc07,scb07} and cluster \cite{Ro09} radioactivities. Folding model potentials using this effective interaction provide excellent descriptions for elastic and inelastic scattering and the nuclear deformation parameters extracted from inelastic scattering of protons \cite{Gu05,Gu06} agree well with other available results. 

    We present a systematic study of the properties of pure hadronic and hybrid compact stars. The nuclear equation of state (EoS) for $\beta$-equilibrated neutron star (NS) matter obtained using density dependent effective nucleon-nucleon interaction satisfies the constraints from the observed flow data from heavy-ion collisions. Depending on model, the energy density of quark matter can be lower than that of this nuclear EoS at higher densities implying the possibility of transition to quark matter inside the core and the transition density depends on the particular quark matter model used. We solve the Einstein's field equations for rotating stars using pure nuclear matter and quark core. The $\beta$- equilibrated neutron star matter, along with different EoSs for crust, is able to describe highly massive compact stars but find that the nuclear to quark matter deconfinement transition inside neutron stars causes reduction in their masses. The recent observations of the binary millisecond pulsar J1614-2230 by P. B. Demorest et al. \cite{De10} and PSR J0348+0432 by J. Antoniadis et al. \cite{An13} suggest that the masses lie within 1.97$\pm$0.04 M$_\odot$ and 2.01$\pm$0.04 M$_\odot$, respectively, where M$_\odot$ is the solar mass. In conformity with recent observations, pure nucleonic EoS determines that the maximum mass of NS rotating with frequency below r-mode instability is $\sim$1.95 M$_\odot$ with radius $\sim$10 kilometers. Although compact stars with quark cores rotating with Kepler's frequency have masses up to $\sim$2 M$_\odot$, but if the maximum frequency is limited by the r-mode instability, the maximum mass of $\sim$1.7 M$_\odot$ turns out to be lower than the observed masses of 1.97$\pm$0.04 M$_\odot$ and 2.01$\pm$0.04 M$_\odot$, by far the highest yet measured with such certainty and thus excluding quark cores within the model we considered for such massive pulsars.
    
\section{Effective nucleon-nucleon interaction and its density dependence from nuclear matter calculations}

    The nuclear matter EoS is calculated using the isoscalar and the isovector \cite{Sa83} components of M3Y interaction along with density dependence. The density dependence of the effective interaction, DDM3Y, is completely determined from nuclear matter calculations. The equilibrium density of the nuclear matter is determined by minimizing the energy per nucleon. The energy variation of the zero range potential is treated accurately by allowing it to vary freely with the kinetic energy part $\epsilon^{kin}$ of the energy per nucleon $\epsilon$ over the entire range of $\epsilon$. This is not only more plausible, but also yields excellent result for the incompressibility $K_\infty$ of the SNM which does not suffer from the superluminosity problem. 
        
    In a Fermi gas model of interacting neutrons and protons, with isospin asymmetry $X= \frac{\rho_n - \rho_p} {\rho_n + \rho_p},~~~~\rho = \rho_n+\rho_p,$ where $\rho_n$, $\rho_p$ and $\rho$ are the neutron, proton and nucleonic densities respectively, the energy per nucleon for isospin asymmetric nuclear matter can be derived as \cite{BCS08}

\begin{equation}
 \epsilon(\rho,X) = [\frac{3\hbar^2k_F^2}{10m}] F(X) + (\frac{\rho J_v C}{2}) (1 - \beta\rho^n)  
\label{seqn1}
\end{equation}
\noindent
where $m$ is the nucleonic mass, $k_F$=$(1.5\pi^2\rho)^{\frac{1}{3}}$ which equals Fermi momentum in case of symmetric nuclear matter (SNM), the kinetic energy per nucleon $\epsilon^{kin}$=$[\frac{3\hbar^2k_F^2}{10m}] F(X)$ with $F(X)$=$[\frac{(1+X)^{5/3} + (1-X)^{5/3}}{2}]$ and $J_v$=$J_{v00} + X^2 J_{v01}$, $J_{v00}$ and $J_{v01}$ represent the volume integrals of the isoscalar and the isovector parts of the M3Y interaction. The isoscalar $t_{00}^{M3Y}$ and the isovector $t_{01}^{M3Y}$ components of M3Y interaction potential are given by $t_{00}^{M3Y}(s, \epsilon)$=7999$\frac{\exp( - 4s)}{4s}$-$2134\frac{\exp( - 2.5s)}{2.5s}$+$J_{00}$(1-$\alpha\epsilon$)$\delta(s)$, $t_{01}^{M3Y}(s, \epsilon)$=-4886$\frac{\exp( - 4s)}{4s}$+$1176\frac{\exp( - 2.5s)}{2.5s}$+$J_{01}$(1-$\alpha\epsilon$)$\delta(s)$ $J_{00}$=-276 MeVfm$^3$, $J_{01}$=228 MeVfm$^3$, $\alpha=0.005$MeV$^{-1}$. The DDM3Y effective NN interaction is given by $v_{0i}(s,\rho, \epsilon) = t_{0i}^{M3Y}(s, \epsilon) g(\rho)$ where the density dependence $g(\rho) = C (1 - \beta \rho^n)$ with $C$ and $\beta$ being the constants of density dependence.

    The Eq.(1) can be differentiated with respect to $\rho$ to yield equation for $X=0$:  
    
\begin{equation}
 \frac{\partial\epsilon}{\partial\rho} = [\frac{\hbar^2k_F^2}{5m\rho}] + \frac{J_{v00} C}{2} [1 - (n+1)\beta\rho^n] 
-\alpha J_{00} C [1 - \beta\rho^n]  [\frac{\hbar^2k_F^2}{10m}].
\label{seqn2}
\end{equation}
\noindent
The equilibrium density of the cold SNM is determined from the saturation condition. Then Eq.(1) and Eq.(2) with the saturation condition $\frac{\partial\epsilon}{\partial\rho} = 0$ at $\rho = \rho_{0}$, $\epsilon = \epsilon_{0}$ can be solved simultaneously for fixed values of the saturation energy per nucleon $\epsilon_0$ and the saturation density $\rho_{0}$ of the cold SNM to obtain the values of $\beta$ and $C$. The constants of density dependence $\beta$ and $C$, thus obtained, are given by 

\begin{equation}
 \beta = \frac{[(1-p)+(q-\frac{3q}{p})]\rho_{0}^{-n}}{[(3n+1)-(n+1)p+(q-\frac{3q}{p})]}
\label{seqn3}
\end{equation} 
\noindent

\begin{equation}
 {\rm where}~~~~p = \frac{[10m\epsilon_0]}{[\hbar^2k_{F_0}^2]},~q=\frac{2\alpha\epsilon_0J_{00}}{J^0_{v00}}
\label{seqn4}
\end{equation} 
\noindent
where $J^0_{v00} = J_{v00}(\epsilon^{kin}_0)$ which means $J_{v00}$ at $\epsilon^{kin}=\epsilon^{kin}_0$, the kinetic energy part of the saturation energy per nucleon of SNM,  $k_{F_0} = [1.5\pi^2\rho_0]^{1/3}$ and 

\begin{equation}
 C = -\frac{[2\hbar^2k_{F_0}^2] }{ 5mJ^0_{v00} \rho_0 [1 - (n+1)\beta\rho_0^n -\frac{q\hbar^2k_{F_0}^2 (1-\beta\rho_0^n)}{10m\epsilon_0}]},
\label{seqn5}
\end{equation} 
\noindent
respectively. It is quite obvious that the constants of density dependence $C$ and $\beta$ obtained by this method depend on the saturation energy per nucleon $\epsilon_0$, the saturation density $\rho_{0}$, the index $n$ of the density dependent part and on the strengths of the M3Y interaction through the volume integral $J^0_{v00}$. 

    The calculations are performed using the values of the saturation density $\rho_0$=0.1533 fm$^{-3}$ \cite{Sa89} and the saturation energy per nucleon $\epsilon_0$=-15.26 MeV \cite{CB06} for the SNM obtained from the co-efficient of the volume term of Bethe-Weizs\"acker mass formula which is evaluated by fitting the recent experimental and estimated atomic mass excesses from Audi-Wapstra-Thibault atomic mass table \cite{Au03} by minimizing the mean square deviation incorporating correction for the electronic binding energy \cite{Lu03}. In a similar recent work, including surface symmetry energy term, Wigner term, shell correction and proton form factor correction to Coulomb energy also, $a_v$ turns out to be 15.4496 MeV \cite{Ro06} ($a_v$ =14.8497 MeV when $A^0$ and $A^{1/3}$ terms are also included). Using the usual values of $\alpha$=0.005 MeV$^{-1}$ for the parameter of energy dependence of the zero range potential and $n$=2/3, the values obtained for the constants of density dependence $C$ and $\beta$ and the SNM incompressibility $K_\infty$ are 2.2497, 1.5934 fm$^2$ and 274.7 MeV, respectively. The saturation energy per nucleon is the volume energy coefficient and the value of -15.26$\pm$0.52 MeV covers, more or less, the entire range of values obtained for $a_v$ for which now the values of $C$=2.2497$\pm$0.0420, $\beta$=1.5934$\pm$0.0085 fm$^2$ and the SNM incompressibility $K_\infty$=274.7$\pm$7.4 MeV.  
    
\begin{figure}[t]
\eject\centerline{\epsfig{file=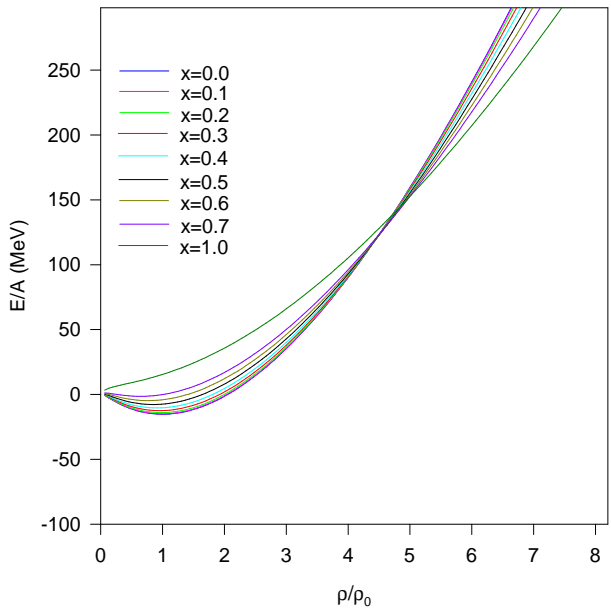,height=8cm,width=8cm}}
\caption
{Plots of energy per nucleon $\epsilon$=E/A of nuclear matter as functions of $\rho/\rho_0$ for different isospin asymmetry $X$.}
\label{fig1}
\vspace{-0.5cm}
\end{figure} 

\vspace{-0.2cm}      
\section{The equation of state}

\subsection{Symmetric and asymmetric nuclear matter}

    The EoSs of the symmetric and the asymmetric nuclear matter describing energy per nucleon as a function of nucleonic density can be obtained by setting $X=0$ and non-zero, respectively, in Eq.(1). The plots of the energy per nucleon $\epsilon$ of nuclear matter with different isospin asymmetry $X$ as functions of $\rho/\rho_0$ for the present calculations are presented in Fig.-1.
    
    The incompressibility or the compression modulus of the SNM, which is a measure of the curvature of an EoS at saturation density and defined as $k_F^2\frac{\partial^2\epsilon}{\partial{k_F^2}} \mid_{k_F=k_{F_0}}$, measures the stiffness of an EoS and obtained theoretically using Eq.(1) for X=0. The incompressibilities for isospin asymmetric nuclear matter are evaluated at saturation densities $\rho_s$ with the condition $\frac{\partial\epsilon}{\partial\rho}=0$ which corresponds to vanishing pressure. The incompressibility $K_s$ for isospin asymmetric nuclear matter is therefore expressed as 

\begin{eqnarray}
 &&K_s = -\frac{3\hbar^2k_{F_s}^2}{5m} F(X) - \frac{9 J^s_v C n(n+1) \beta\rho_s^{n+1}}{2} \nonumber \\
 &&- 9\alpha J C [1-(n+1)\beta\rho_s^n] [\frac{\rho_s\hbar^2k_{F_s}^2}{5m}] F(X) \nonumber \\
 &&+ [\frac{3\rho_s\alpha J C (1-\beta\rho_s^n)\hbar^2k_{F_s}^2}{10m}] F(X). 
\label{seqn6}
\end{eqnarray} 
\noindent
Here $k_{F_s}$ means that the $k_F$ is evaluated at saturation density $\rho_s$. $J^s_v=J^s_{v00} + X^2 J^s_{v01}$ is the $J_v$ at $\epsilon^{kin}=\epsilon^{kin}_s$ which is the kinetic energy part of the saturation energy per nucleon $\epsilon_s$ and $J=J_{00} + X^2 J_{01}$.

    In Table-I incompressibility of isospin asymmetric nuclear matter  $K_s$ as a function of the isospin asymmetry parameter $X$, using the usual value of $n$=2/3 and energy dependence parameter $\alpha$=0.005 MeV$^{-1}$, is provided. The magnitude of the incompressibility $K_s$ decreases with the isospin asymmetry $X$ due to lowering of the saturation densities $\rho_s$ with $X$ as well as decrease in the EoS curvature. At high isospin asymmetry $X$, the isospin asymmetric nuclear matter does not have a minimum signifying that it can never be bound by itself due to nuclear interaction. However, the $\beta$ equilibrated nuclear matter which is a highly neutron rich asymmetric nuclear matter exists in the core of the neutron stars since its E/A is lower than that of SNM at high densities and is unbound by the nuclear force but can be bound due to high gravitational field realizable inside neutron stars.      

\noindent 
\begin{table}
\centering
\caption{Incompressibility of isospin asymmetric nuclear matter using the usual value of $n$=2/3 and energy dependence  parameter $\alpha$=0.005 MeV$^{-1}$.}
\begin{tabular}{ccc}
\hline
\hline
$X$&$\rho_s$& $K_s$      \\
\hline
 & fm$^{-3}$ &MeV    \\ 
\hline
 0.0&0.1533&274.7 \\ 
 0.1&0.1525&270.4 \\ 
 0.2&0.1500&257.7 \\ 
 0.3&0.1457&236.6 \\ 
 0.4&0.1392&207.6 \\ 
 0.5&0.1300&171.2 \\  \hline
\hline
\label{table1}
\end{tabular} 
\end{table}
\nopagebreak

    The pressure P of SNM and pure neutron matter (PNM) are plotted in Fig.-2 and Fig.-3 respectively as functions of $\rho/\rho_0$. The continuous lines represent the present calculations whereas the dotted lines represent the same using the A18 model using variational chain summation (VCS) of Akmal et al. \cite{Ak98} for the SNM and PNM. The dash-dotted line of Fig.-2 represents plot of $P$ versus $\rho/\rho_0$ for SNM for RMF using NL3 parameter set \cite{La97} and the area enclosed by the continuous line corresponds to the region of pressures consistent with the experimental flow data \cite{Da02}. It is interesting to note that the RMF-NL3 incompressibility for SNM is 271.76 MeV \cite{La99} which is about the same as 274.7$\pm$7.4 MeV obtained for the present calculation. The areas enclosed by the continuous and the dashed lines in Fig.-3 correspond to the pressure regions for neutron matter consistent with the experimental flow data after inclusion of the pressures from asymmetry terms with weak (soft NM) and strong (stiff NM) density dependences, respectively \cite{Da02}. Although, the parameters of the density dependence of DDM3Y interaction have been tuned to reproduce $\rho_0$ and $\epsilon_0$ which are obtained from finite nuclei, the agreement of the present EoS with the experimental flow data, where the high density behaviour looks phenomenologically confirmed, justifies its extrapolation to high density.

\begin{figure}[t]
\eject\centerline{\epsfig{file=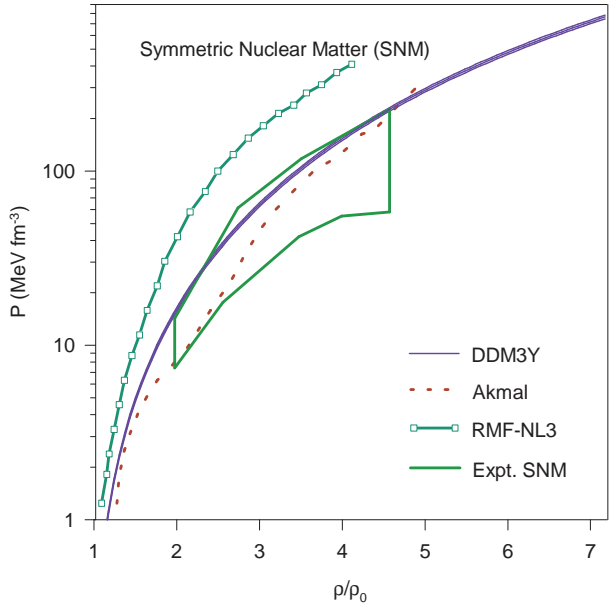,height=8cm,width=8cm}}
\caption
{The pressure P of SNM (spin and isospin symmetric nuclear matter) as a function of $\rho/\rho_0$. The continuous lines represent the present calculations using $\epsilon_0$ = -15.26$\pm$0.52 MeV. The dotted line represents the same using the A18 model using variational chain summation (VCS) of Akmal et al. \cite{Ak98}, the dash-dotted line represents the RMF calculations using NL3 parameter set \cite{La97} whereas the area enclosed by the continuous line corresponds
to the region of pressures consistent with the experimental flow data \cite{Da02} for SNM.}
\label{fig2}
\end{figure}

    The present status of experimental determination of the SNM incompressibility from the compression modes ISGMR and isoscalar giant dipole resonance (ISGDR) of nuclei infers \cite{Sh06} that due to violations of self consistency in HF-RPA calculations of the strength functions of giant resonances result in shifts in the calculated values of the centroid energies which may be larger in magnitude than the current experimental uncertainties. In fact, the prediction of $K_\infty$ lying in the range of 210-220 MeV were due to the use of a not fully self-consistent Skyrme calculations \cite{Sh06}. Correcting for this drawback, Skyrme parmetrizations of SLy4 type predict $K_\infty$ values in the range of 230-240 MeV \cite{Sh06}. Moreover, it is possible to build bona fide Skyrme forces so that the SNM incompressibility is close to the relativistic value, namely 250-270 MeV. Hence, from the ISGMR experimental data, conclusion may be drawn that $K_\infty \approx$ 240 $\pm$ 20 MeV.

\begin{figure}[t]
\eject\centerline{\epsfig{file=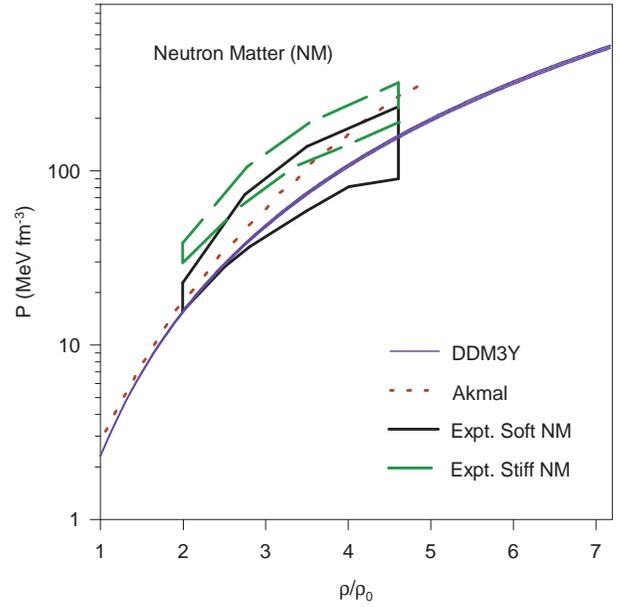,height=8cm,width=8cm}}
\caption
{The pressure P of PNM (pure neutron matter) as a function of $\rho/\rho_0$. The continuous lines represent the present calculations using $\epsilon_0$ = -15.26$\pm$0.52 MeV. The dotted line represents the same using the A18 model using variational chain summation (VCS) of Akmal et al. \cite{Ak98} whereas the areas enclosed by the continuous and the dashed lines correspond to the pressure regions for neutron matter consistent with the experimental flow data after inclusion of the pressures from asymmetry terms with weak (soft NM) and strong (stiff NM) density dependences, respectively \cite{Da02}.}
\label{fig3}
\vspace{-0.55cm}
\end{figure}
   
    The ISGDR data tend to point to lower values \cite{Lu04,Yo04} for $K_\infty$. However, there is consensus that the extraction of $K_\infty$ is in this case more problematic for various reasons. In particular, the maximum cross-section for ISGDR decreases very strongly at high excitation energy and may drop below the current experimental sensitivity for excitation energies \cite{Sh06} above 30 and 26 MeV for $^{116}$Sn and $^{208}$Pb, respectively. The present non-relativistic mean field model estimate for the nuclear incompressibility $K_\infty$ for SNM using DDM3Y interaction is rather close to the theoretical estimates obtained using relativistic mean field models and close to the upper limit of the recent values \cite{Yo05} extracted from experiments. The results of microscopic calculations, with Gogny effective interactions \cite{Bl80} which include nuclei where pairing correlations are important, reproduce experimental data on heavy nuclei, for a compression modulus in the range about 220 MeV \cite{Bl95}. The recent acceptable value \cite{Vr03,Sh09} of SNM incompressibility lies in the range of 250-270 MeV and calculated value of 274.7$\pm$7.4 MeV is a good theoretical result and is only slightly too high.       
 
\subsection{Incompressibility, isobaric incompressibility, symmetry energy and its slope}

    The EOS of isospin asymmetric nuclear matter, given by Eq.(1) can be expanded as

\begin{equation}
  \epsilon(\rho,X) =  \epsilon(\rho,0) + E_{sym}(\rho) X^2 + O ( X^4)
\label{seqn7}
\end{equation} 
\noindent
in general and $E_{sym}(\rho)= \frac{1}{2} \frac{\partial^2\epsilon(\rho,X)}{\partial{X^2}} \mid_{X=0}$ is termed as the NSE. The absence of odd-order terms in $X$ in Eq.(7) is due to the exchange symmetry between protons and neutrons in nuclear matter when one neglects the Coulomb interaction and assumes the charge symmetry of nuclear forces. The higher-order terms in $X$ are negligible and to a good approximation, the density-dependent NSE $E_{sym}(\rho)$ can be extracted using following equation \cite{Kl06}

\begin{equation}
 E_{sym}(\rho)=\epsilon(\rho,1) -\epsilon(\rho,0)
\label{seqn8}
\end{equation}
\noindent
which can be obtained using Eq.(1) and represents a penalty levied on the system as it departs from the symmetric limit of equal number of protons and neutrons and can be defined as the energy required per nucleon to change the SNM to pure neutron matter (PNM). In Fig.-4 the NSE is plotted as a function of $\rho/\rho_0$ for the present calculation using DDM3Y interaction, and compared with those for Akmal-Pandharipande-Ravenhall \cite{Ak98} and MDI interactions \cite{Zh09}. 

\begin{figure}[t]
\vspace{-0.0cm}
\eject\centerline{\epsfig{file=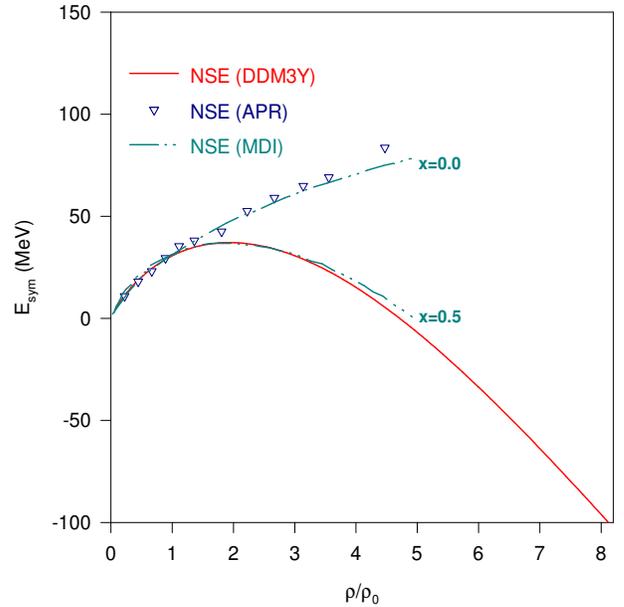,height=8cm,width=8cm}}
\caption
{The NSE (nuclear symmetry energy $E_{sym}$) is plotted as a function of $\rho/\rho_0$ for the present calculation using DDM3Y interaction and its comparison, with those for Akmal-Pandharipande-Ravenhall (APR) \cite{Ak98} and MDI interactions for the variable x=0.0, 0.5 defined in Ref. \cite{Zh09}.}
\label{fig4}
\vspace{-0.0cm}
\end{figure}

    The volume symmetry energy coefficient $S_v$ extracted from nuclear masses provides a constraint on the NSE at nuclear density $E_{sym}(\rho_0)$. The value of $S_v$=30.048 $\pm$0.004 MeV extracted \cite{Mu07} from the measured atomic mass excesses of 2228 nuclei is reasonably close to the theoretical estimate of the value of NSE at saturation density $E_{sym}(\rho_0)$=30.71$\pm$0.26 MeV obtained from the present calculations using DDM3Y interaction. If one uses the alternative definition of $E_{sym}(\rho)= \frac{1}{2} \frac{\partial^2\epsilon(\rho,X)}{\partial{X^2}} \mid_{X=0}$, the value of NSE at the saturation density remains almost the same which is 30.03$\pm$0.26 MeV. Empirically the value of $E_{sym}(\rho_0)\approx$ 30 MeV \cite{Da03,St05,Po03} seems well established. Theoretically different parametrizations of the relativistic mean-field (RMF) models, which fit observables for isospin symmetric nuclei well, lead to a relatively wide range of predictions of 24-40 MeV for $E_{sym}(\rho_0)$. The present result of 30.71$\pm$0.26 MeV is close to that using Skyrme interaction SkMP (29.9 MeV) \cite{Be89}, Av18+$\delta v$+UIX$^*$ variational calculation (30.1 MeV) \cite{Ak98}. 
 
    Around nuclear matter saturation density $\rho_0$, the NSE $E_{sym}(\rho)$ can be expanded to second order in density as 

\begin{equation}
 E_{sym}(\rho)= E_{sym}(\rho_0) + \frac{L}{3} {\Big (}\frac{\rho - \rho_0}{\rho_0}{\Big )}+ \frac{K_{sym}}{18}{\Big (}\frac{\rho - \rho_0}{\rho_0}{\Big )}^2
\label{seqn9}
\end{equation}
\noindent
where $L$ and $K_{sym}$ represents the slope and curvature parameters of NSE at $\rho_0$ and hence $L= 3\rho_0 \frac{\partial E_{sym}(\rho)}{\partial\rho} \mid_{\rho=\rho_0}$ and $K_{sym}= 9\rho_0^2 \frac{\partial^2 E_{sym}(\rho)}{\partial {\rho^2}} \mid_{\rho=\rho_0}$. The $L$ and $K_{sym}$ characterize the density dependence of the NSE around normal nuclear matter density and thus carry important information on the properties of NSE at both high and low densities. In particular, the slope parameter $L$ has been found to correlate linearly with the neutron-skin thickness of heavy nuclei and can be determined from the measured thickness of neutron skin of such nuclei \cite{Pi05,Ch05,SL05}. Recently, this has been possible \cite{Ce09} although there are large uncertainties in the experimental measurements.

    The Eq.(8) can be differentiated twice with respect to nucleonic density $\rho$ using Eq.(1) to yield  

\begin{eqnarray}
 &&\frac{\partial E_{sym}}{\partial \rho} =\frac{2}{5}(2^{2/3}-1)\frac{E^0_F}{\rho}(\frac{\rho}{\rho_0})^{2/3}+\frac{C}{2}[1-(n+1)\beta\rho^n]  \nonumber \\
&&\times J_{v01}(\epsilon^{kin}_{X=1})-\frac{\alpha J_{01}C}{5}E^0_F(\frac{\rho}{\rho_0})^{2/3} [1-\beta\rho^n]F(1) \nonumber \\
&&-(2^{2/3}-1)\frac{\alpha J_{00}C}{5}E^0_F(\frac{\rho}{\rho_0})^{2/3}[1-\beta\rho^n] \nonumber \\ &&-\frac{3}{10}(2^{2/3}-1)\alpha J_{00}CE^0_F(\frac{\rho}{\rho_0})^{2/3}[1-(n+1)\beta\rho^n]
\label{seqn10}
\end{eqnarray} 
\noindent

\begin{eqnarray}
&&\frac{\partial^2 E_{sym}}{\partial \rho^2} =-\frac{2}{15}(2^{2/3}-1)\frac{E^0_F}{\rho^2}(\frac{\rho}{\rho_0})^{2/3}-\frac{C}{2}n(n+1)\beta\rho^{n-1}\nonumber \\
&&\times J_{v01} (\epsilon^{kin}_{X=1}) -\frac{2\alpha J_{01}C}{5}\frac{E^0_F}{\rho}(\frac{\rho}{\rho_0})^{2/3}[1-(n+1)\beta\rho^n]F(1)  
 \nonumber \\
&&+\frac{\alpha J_{01}C}{15}\frac{E^0_F}{\rho}(\frac{\rho}{\rho_0})^{2/3}[1-\beta\rho^n]F(1) \nonumber \\
&&+(2^{2/3}-1)\frac{\alpha J_{00}C}{15}\frac{E^0_F}{\rho}(\frac{\rho}{\rho_0})^{2/3} [1-\beta\rho^n]  \nonumber \\
&&-\frac{2}{5}(2^{2/3}-1)\alpha J_{00}C\frac{E^0_F}{\rho}(\frac{\rho}{\rho_0})^{2/3} [1-(n+1) \beta\rho^n] \nonumber \\ 
&&+\frac{3}{10}(2^{2/3}-1)\alpha J_{00}CE^0_F(\frac{\rho}{\rho_0})^{2/3}n(n+1)\beta\rho^{n-1}
\label{seqn11}
\end{eqnarray} 
\noindent
where the Fermi energy $E^0_F=\frac{\hbar^2k_{F_0}^2}{2m}$ for the SNM at ground state. Eqs.(10,11) at $\rho$=$\rho_0$ are used to evaluate the values of $L$ and $K_{sym}$. 

    The isobaric incompressibility for infinite nuclear matter can be expanded in the power series of isospin asymmetry $X$ as $K_\infty (X) = K_\infty + K_\tau X^2 + K_4 X^4 + O(X^6)$. The magnitude of the higher-order $K_4$ parameter is generally quite small compared to $K_\tau$ \cite{Ch09}. The latter essentially characterizes the isospin dependence of the incompressibility at saturation density and can be expressed as $K_\tau=K_{sym}-6L-\frac{Q_0}{K_\infty}L=K_{asy}-\frac{Q_0}{K_\infty}L$ where $Q_0$ is the third-order derivative parameter of symmetric nuclear matter at $\rho_0$ given by 
    
\begin{equation}
Q_0 = 27\rho_0^3 \frac{\partial^3 \epsilon(\rho,0)}{\partial {\rho^3}} \mid_{\rho=\rho_0}. 
\label{seqn12}
\end{equation}
\noindent
Using Eq.(1) one obtains

\begin{eqnarray}
&&\frac{\partial^3 \epsilon(\rho,X)}{\partial {\rho^3}} 
=-\frac{CJ_v(\epsilon^{kin})n(n+1)(n-1)\beta\rho^{n-2}}{2}  \nonumber \\ &&+\frac{8}{45}\frac{E^0_F}{\rho^3}F(X)(\frac{\rho}{\rho_0})^\frac{2}{3} 
+\frac{3\alpha JC}{5}n(n+1)\beta\rho^{n-1}\frac{E^0_F}{\rho} \nonumber \\ 
&&\times F(X)(\frac{\rho}{\rho_0})^\frac{2}{3} 
+\frac{\alpha JC}{5}[1-(n+1)\beta\rho^n] \frac{E^0_F}{\rho^2}F(X)(\frac{\rho}{\rho_0})^\frac{2}{3} \nonumber \\
&&-\frac{4\alpha JC}{45}[1-\beta\rho^n]\frac{E^0_F}{\rho^2}F(X)(\frac{\rho}{\rho_0})^\frac{2}{3}
\label{seqn13}
\end{eqnarray}
\noindent
where the Fermi energy $E^0_F$=$\frac{\hbar^2k_{F_0}^2}{2m}$ for the SNM at ground state,  $k_{F_0}$=$(1.5\pi^2\rho_0)^{\frac{1}{3}}$ and $J$=$J_{00}$+$X^2J_{01}$. Thus 

\begin{eqnarray}
&&\frac{\partial^3 \epsilon(\rho,0)}{\partial {\rho^3}} \mid_{\rho=\rho_0}
=-\frac{CJ_{v00}(\epsilon_0^{kin})n(n+1)(n-1)\beta\rho_0^{n-2}}{2} \nonumber \\  
&&+\frac{8}{45}\frac{E^0_F}{\rho_0^3}+\frac{3\alpha J_{00}C}{5}n(n+1)\beta\rho_0^{n-1} \frac{E^0_F}{\rho_0}+\frac{\alpha J_{00}C}{5} \nonumber \\   
&&\times [1-(n+1)\beta\rho_0^n]\frac{E^0_F}{\rho_0^2}-\frac{4\alpha J_{00}C}{45}[1-\beta\rho_0^n]\frac{E^0_F}{\rho_0^2}
\label{seqn14}
\end{eqnarray}
\noindent
where $\epsilon_0^{kin}$ is the kinetic energy part of the saturation energy per nucleon $\epsilon_0$. The calculations are performed using the values of the saturation density $\rho_0$=0.1533 fm$^{-3}$, the saturation energy per nucleon $\epsilon_0=-15.26\pm0.52$ MeV for the SNM and $n=\frac{2}{3}$ \cite{CBS09}. The saturation energy per nucleon is the volume energy coefficient $a_v$ of liquid drop model and the value of -15.26$\pm$0.52 MeV covers, more or less, the entire range of values obtained for $a_v$ for which the values of $C$ and $\beta$ are 2.2497$\pm$0.0420 and 1.5934$\pm$0.0085 fm$^2$, respectively \cite{BCS08}. Collisions involving $^{112}$Sn and $^{124}$Sn nuclei can be simulated with the improved quantum molecular dynamics transport model to reproduce isospin diffusion data from two different observables and the ratios of neutron and proton spectra. Constraints on the density dependence of the symmetry energy at subnormal density can be obtained \cite{Ts09} by comparing these data to calculations performed over a range of symmetry energies at saturation density and different representations of the density dependence of the symmetry energy. The results of the present calculations for $K_\infty$, $L$, $E_{sym}(\rho_0)$ and density dependence of $E_{sym}(\rho)$ \cite{CBS09} are consistent with these constraints \cite{Ts09}. In Table-II, the values of $K_\infty$, $E_{sym}(\rho_0)$, $L$, $K_{sym}$ and $K_\tau$ are listed and compared with the corresponding quantities obtained with relativistic mean field (RMF) models \cite{Pi09}. 

\begin{table*}[htbp]
\centering
\caption{Results of the present calculations (DDM3Y) of incompressibility of isospin symmetric nuclear matter $K_\infty$, nuclear symmetry energy at saturation density $E_{sym}(\rho_0)$, the slope $L$ and the curvature $K_{sym}$ parameters of the nuclear symmetry energy, the approximate isospin dependent part $K_{asy}$ and the exact part $K_\tau$ of the isobaric incompressibility (all in MeV) are compared with those obtained with RMF models \cite{Pi09}.}
\begin{tabular}{cccccccc}
\hline
\hline
Model&$K_\infty$&$E_{sym}(\rho_0)$&$L$&$K_{sym}$&$K_{asy}$&$Q_0$&$K_\tau$\\ 
\hline
 This work &$274.7\pm7.4$&$30.71\pm0.26$&$45.11\pm0.02$&$-183.7\pm3.6$&$-454.4\pm3.5$&$-276.5\pm10.5$&$-408.97\pm3.01$ \\ 
 FSUGold&230.0&32.59&60.5&-51.3&-414.3&-523.4&-276.77\\
 NL3&271.5&37.29&118.2&+100.9&-608.3&+204.2&-697.36 \\
 Hybrid&230.0&37.30&118.6&+110.9&-600.7&-71.5&-563.86\\ 
\hline
\hline
\label{table2}
\end{tabular} 
\end{table*}
\noindent 

\begin{figure}[t]
\vspace{0.0cm}
\eject\centerline{\epsfig{file=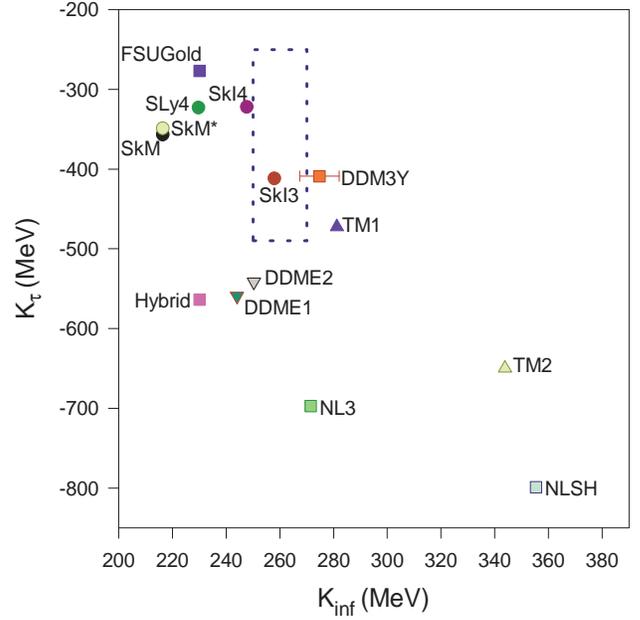,height=8.2cm,width=8.2cm}}
\caption
{The $K_\tau$ is plotted against  $K_\infty$ ($K_{inf}$) for the present calculation using DDM3Y interaction and compared with other predictions as tabulated in Refs. \cite{Pi09,Sa07}. The dotted rectangular region encompasses the values of $K_\infty=250-270$ MeV \cite{Sh09} and $K_\tau=-370\pm120$ MeV \cite{Ch09}.}
\label{fig5}
\vspace{0.0cm}
\end{figure}

    There seems to remain controversy over what is a reasonable value of incompressibility \cite{Sh06}. In the following we do not justify any particular value for $K_\infty$ but present our results in the backdrop of others for an objective view of the current scenario which, we stress, is still evolving. In Fig.-5, $K_\tau$ is plotted against $K_\infty$ for the present calculation using DDM3Y interaction and compared with the predictions of FSUGold, NL3, Hybrid \cite{Pi09}, SkI3, SkI4, SLy4, SkM, SkM*, NLSH, TM1, TM2, DDME1 and DDME2 as given in Table-I of Ref. \cite{Sa07}. The dotted rectangular region encompasses the recent values of $K_\infty=250-270$ MeV \cite{Sh09} and $K_\tau=-370\pm120$ MeV \cite{Ch09}. Although both DDM3Y and SkI3 are within the above region, unlike DDM3Y the $L$ value for SkI3 is 100.49 MeV which is much above the acceptable limit of 45-75 MeV \cite{Wa09} whereas DDME2 which gives $L=51$ MeV is reasonably close to the rectangular region. It is worthwhile to mention here that the DDM3Y interaction with the same ranges, strengths and density dependence which gives $L=45.11\pm0.02$ here, provides good descriptions of scattering (elastic and inelastic), proton radioactivity \cite{BCS08} and $\alpha$ radioactivity of superheavy elements \cite{CSB06,scb07}. The present NSE is `super-soft' because it increases initially with nucleonic density up to about two times the normal nuclear density and then decreases monotonically (hence `soft') and becomes negative (hence `super-soft') at higher densities (about 4.7 times the normal nuclear density) \cite{BCS08,CBS09} and is consistent with the recent evidence for a soft NSE at suprasaturation densities \cite{Zh09} and with the fact that the super-soft nuclear symmetry energy preferred by the FOPI/GSI experimental data on the $\pi^+/\pi^-$ ratio in relativistic heavy-ion reactions can readily keep neutron stars stable if the non-Newtonian gravity proposed in the grand unification theories is considered \cite{We09}. 

\section{Nuclear Scattering}

\subsection{Elastic scattering using potentials from folding effective interaction}

    The microscopic proton-nucleus interaction potentials are obtained by single folding the density distribution of the nucleus with M3Y effective interaction along with the density dependence as

\begin{equation}
V_N(R) = \int \rho(\vec{r}) v_{00}(|\vec{R}-\vec{r}|)d^3\vec{r} \\
\label{seqn15}
\end{equation}
where $\rho(\vec{r})$ is density of the nucleus at $\vec{r}$ and $v_{00}$ is the effective interaction between two nucleons at the sites $\vec{R}$ and $\vec{r}$. The parameters of the density dependence, $C$=2.2497 and $\beta$=1.5934 fm$^2$, obtained here from the nuclear matter calculations assuming kinetic energy dependence of zero range potential, are used. The nuclear ground state densities have been calculated in the framework of spherical Hartree Fock plus BCS calculations in co-ordinate space using two different parameter sets of the Skyrme interactions. There are negligible differences in the ground state densities of the nuclei with the two different parameter sets (for SIII~\cite{BE75} and SkM*~\cite{BA82}). We use SkM* parameterization for the $^{18}$O ground state density and used it for calculating the single folding potential and form factor. We carried out the same procedure for $^{18}$Ne and for SkM* the binding energies per nucleon are in good agreement (within 0.1 MeV) with the experimental values for  both the nuclei.

    The phenomenological optical potentials have the form 
    
\begin{eqnarray}
V_{\rm pheno}(r) = && -V_o~f_o(r)-iW_vf_v(r) + 4ia_sW_s(d/dr) f_s(r)  \nonumber \\
+ 2(\hbar/m_{\pi}c)^2 && V_{s.o} [1/r] (d/dr) f_{s.o}(r)({\bf L.S})+ V_{\rm coul}
\label{seqn16}
\end{eqnarray}
where, $f_x(r)~= [1~+~exp(\frac{r-R_x}{a_x})]^{-1}$, $R_{x}~=~r_{x}A^{1/3}$ and $x~=~o,v,s,s.o$. The subscripts $o,v,s,s.o$ denote real, volume imaginary, surface imaginary and spin-orbit respectively and  $V_o$, $W_v$ ($W_s$) and $V_{s.o}$ are the strengths of the real, volume (surface) imaginary and spin-orbit potentials respectively. $V_{\rm coul}$ is the Coulomb potential of a uniformly charged sphere of radius 1.20~$A^{1/3}$.

    In semi-microscopic analysis both the volume real ($V$) and volume imaginary ($W$) parts of the potentials (generated microscopically by folding model) are assumed to have the same shape, i.e. $V_{\rm micro}(r)~=~V + iW$ = ($N_{\rm R}$ + $iN_{\rm I}$)$V_N$($r$) where, $N_{\rm R}$ and $N_{\rm I}$ are the renormalization factors for real and imaginary parts respectively~\cite{SA97}. Thus the potentials for elastic scattering analysis include real and volume imaginary terms (folded potentials) and also surface imaginary and spin-orbit terms (best fit phenomenological potentials).

    For each angular distribution, best fits are obtained by minimizing $\chi^2$/N, where $\chi^2$ = $\sum_{k = 1}^{\rm N} \left[\frac{\sigma_{th}(\theta_k)~-\sigma_{ex}(\theta_k)}{\Delta\sigma_{ex} (\theta_k)}\right]^2$, where $\sigma_{th}$ and $\sigma_{ex}$ are the theoretical and experimental cross sections respectively, at angle $\theta_k$, $\Delta\sigma_{ex}$ is the experimental error and N is
the number of data points. 

\subsection{Inelastic scattering and nuclear deformation parameter} 

    The radius $R$ of the deformed target nucleus can be expanded as

\begin{equation}
 R = R_0 [1 + \beta_2 Y^0_2 ( \theta, \phi ) ] 
\label{seqn17}
\end{equation}                                                                                               \noindent     
for l=2 transition, where $Y^0_2$ is the corresponding spherical harmonic, $\beta_2$ is the deformation parameter and $R_0$ ($\approx r_{\rm rms}^V$) is the radius of the spherical nucleus having the same volume as that of the deformed target nucleus. Hence the deformed potential $ V(r,R)$ can be expanded in a Taylor series about $R_0$ in terms of the calculated spherically symmetric potential $ V_0(r,R_0)$ as following

\begin{equation}
 V(r,R) = V_0( r,R_0 ) - \beta_2 R_0  Y^0_2 ( \theta, \phi ) \frac{dV_0}{dr}
\label{seqn18}
\end{equation}                                                                                               \noindent     
where the first term is the usual spherical folded potential and the second term is responsible for the coupling between elastic and inelastic channels. The T-matrix for the inelastic scattering can therefore be given by 

\begin{eqnarray}
 T_{\beta\alpha}  &&= < \chi_{\beta}\phi_{\beta}| V-V_0 | \chi_{\alpha}\phi_{\alpha}> \nonumber \\
 &&= < \chi_{\beta}\phi_{\beta}| -\beta_2 R_0  Y^0_2 ( \theta, \phi ) dV_0/dr | \chi_{\alpha}\phi_{\alpha}>
\label{seqn19}
\end{eqnarray}                                                                                               \noindent     
where the subscripts $\alpha$ and $\beta$ represent the elastic and inelastic channels respectively. The deformation parameters $\beta_2$ can now be determined by fitting the inelastic scattering angular distribution. 

    The potentials for elastic scattering analysis are subsequently used in the DWBA calculations of inelastic scattering with transferred angular momentum $l$. The calculations are performed using the code DWUCK4~\cite{DWUCK4}. The derivative of the potentials ($\delta \frac{dV}{dr}$) are used as the form factors. The microscopic real and imaginary form factors have the same shape with strengths $N_{\rm R}^{\rm FF}$ and $N_{\rm I}^{\rm FF}$ respectively, where $N_{\rm R,I}^{\rm FF}$ = $N_{\rm R,I}r_{\rm
rms}^V$, where the radius parameter $r_{\rm rms}^V$ is the rms radius of the folded potential. In addition, form factors derived from phenomenological surface imaginary and spin-orbit potentials are included. Obviously, the folded potentials for real and volume imaginary terms and the best fit phenomenological surface imaginary and spin-orbit terms need not obey the local equivalent dispersion relation. The deformation parameters $\delta$ are determined by fitting the inelastic scattering angular distribution. The
renormalizations required for the potentials are reminiscent of those for deuteron and $^6$Li scattering and it may be surmised that it is  for the same reasons; weak binding and ease of breakup and other reaction channels. 

\begin{figure}[t]
\vspace{-1.25cm}
\eject\centerline{\epsfig{file=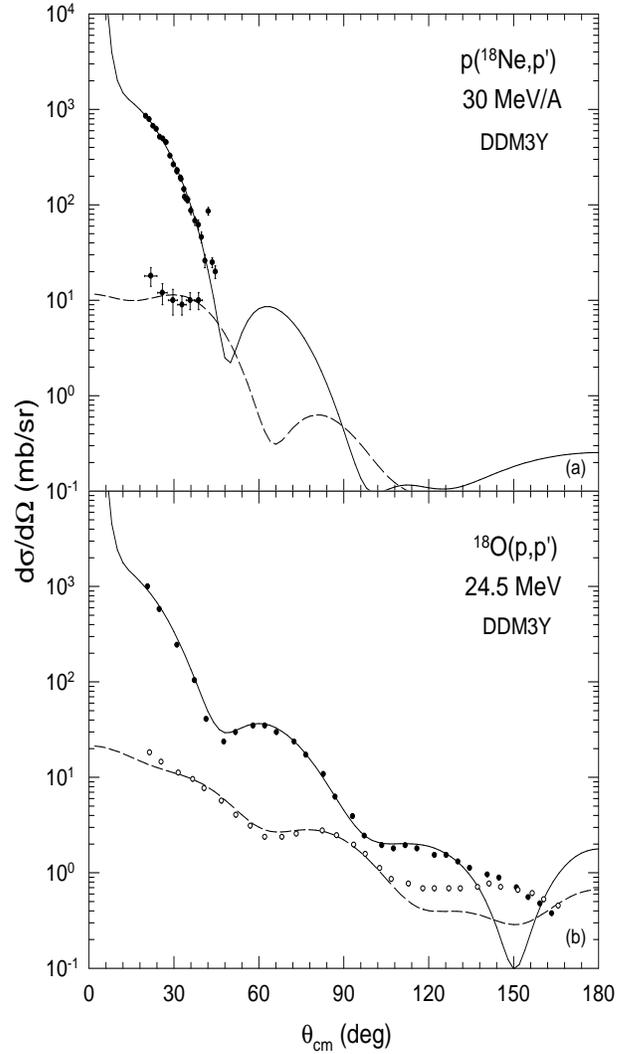,height=14cm,width=8cm}}
\caption
{The experimental angular distributions and folding model calculations of (a) p + $^{18}$Ne at 30A MeV for elastic and inelastic [$E^*$ = 1.890 MeV (2$^+$)] scattering \cite{RI99}, (b) p + $^{18}$O at 24.5A MeV for elastic and inelastic [$E^*$ = 1.982 MeV (2$^+$)] scattering \cite{ES74}. The continuous and  dashed lines correspond to calculations for elastic and inelastic cross sections respectively.}
\label{fig6}
\vspace{0.0cm}
\end{figure}
 
\noindent 
\begin{table}
\centering
\caption{Comparison of nuclear deformation parameters $\delta$ extracted from inelastic scattering and from B(E2) values along with corresponding intrinsic quadrupole moments (in millibarns).}
\begin{tabular}{cccc}
\hline
\hline
Nucleus& $\delta$   &  $\delta$    & Quadrupole \\ \hline
       &Present work& From B(E2) & moment(mb)\\
\hline
 O$^{18}$ &0.33&0.355(8)* &212.9(47)*\\ 
 O$^{20}$ &0.46&0.261(9)*[0.50(4)**]&168(6)* \\ 
 O$^{22}$ &0.26&0.208(41)*&143(28)* \\ 
Ne$^{18}$ &0.40&0.694(34)*&519(25)* \\  \hline
\hline
&*From Ref.\cite{Ra01} &**From Ref.\cite{Je99} &
\label{table3}
\end{tabular}
\end{table}
\nopagebreak
    
    The relationship between the reduced electric quadrupole transition rate $B(E2)$ for the ground state to the $2^+$ state in units of $e^2fm^4$ and the quadrupole deformation parameter $\delta$ is given by \cite{Ra01}

\begin{equation}
\delta(1+0.16\delta+0.20\delta^2+....)= 4 \pi B(E2)^{1/2}/(3ZR^2)
\label{seqn20}
\end{equation}
\noindent 
where $R=1.2A^{1/3}$ fm and  $Z$ is the atomic number. The quadrupole deformations are listed in reference~ \cite{Ra01,Je99}. As can be seen from Table-III, the quadrupole deformation obtained from the present analysis for $^{18,20,22}$O is in excellent agreement while that for $^{18}$Ne is significantly underestimated due to lack of experimental data at forward angles. The inelastic scattering is more sensitive at forward angles due to its relative purity compared to data corresponding to backward angular range where other non-elastic processes also contribute. The quality of the elastic as well as the inelastic fits deteriorates at backward angles and similar deterioration of fits are also seen for proton scattering from other nuclei~\cite{Sa79,KH02}. The reason for this is probably that the full cross section is ascribed to potential scattering while quasi-compound-nucleus formation feeds back into the elastic channel and whose energy-dependence is controlled by barrier-top resonances. The backward angular range classically corresponds to smaller impact parameters. This fact suggests higher compound nuclear formation probabilities at backward angles while those at forward angles are expected to be negligibly small. Since relative contributions of compound elastic and direct elastic are not disentangled, $\chi^2$ was calculated only upto about 90$^o$ in center of mass during fitting of the data. In the present study we find that the realistic DDM3Y effective interaction obtained from sophisticated G-matrix calculations, whose density dependence is determined from nuclear matter calculations, provides good descriptions for the elastic and inelastic scattering of protons (Fig.-6) from the nuclei $^{18}$Ne and $^{18,20,22}$O and the values of the deformation parameters extracted from the calculations for these nuclei are reasonable. 
 
\section{Nuclear decays}
   
\subsection{Proton radioactivity}

    The half lives of the decays of spherical nuclei away from proton drip line by proton emissions are estimated theoretically. The half life of a parent nucleus decaying via proton emission is calculated using the WKB barrier penetration probability. The WKB method is found quite satisfactory and even better than the S-matrix method for calculating half widths of the $\alpha$ decay of superheavy elements \cite{Ma06}. For the present calculations, the zero point vibration energies used here are given by Eq.(5) of ref. \cite{Po86} extended to protons and the experimental $Q$ values \cite{So02} are used. Spherical charge distributions are used for Coulomb interaction potentials. The isovector or the symmetry component \cite{Sa83} of the DDM3Y folded potential $V^{Lane}_N(R)$ has been added to the isoscalar part of the folded potential given by Eq.(15). The nuclear potential $V_N(R)$ of Eq.(15), therefore, has been replaced by $V_N(R)+V^{Lane}_N(R)$ where

\begin{eqnarray}
 V^{Lane}_N(R) =&& \int \int [\rho_{1n}(\vec{r_1})-\rho_{1p}(\vec{r_1})] [\rho_{2n}(\vec{r_2})-\rho_{2p}(\vec{r_2})] \nonumber \\
 && \times v_{01}[|\vec{r_2} - \vec{r_1} + \vec{R}|] d^3r_1 d^3r_2 
\label{seqn21}
\end{eqnarray}
\noindent
where the subscripts 1 and 2 denote the daughter and the emitted nuclei respectively while the subscripts n and p denote neutron and proton densities respectively. With simple assumption that $\rho_{1p}=[\frac{Z_d}{A_d}]\rho$ and $\rho_{1n}=[\frac{(A_d-Z_d)}{A_d}]\rho$, and for the emitted particle being proton $\rho_{2n}(\vec{r_2})- \rho_{2p} (\vec{r_2})=-\rho_2(\vec{r_2})=-\delta(\vec{r_2})$, the Lane potential becomes $ V^{Lane}_N(R) = -[\frac{(A_d-2Z_d)}{A_d}] \int \rho (\vec{r}) v_{01} [|\vec{R} - \vec{r}|] d^3r $ where $v_{01}(s)=t_{01}^{M3Y}(s,E)g(\rho)$ and $A_d$ and $Z_d$ are, respectively, the mass number and the charge number of the daughter nucleus. The inclusion of this Lane potential causes insignificant changes in the lifetimes. The same set of data of ref. \cite{Bal05} has been used for the present calculations using $C$=2.2497 and $\beta$=1.5934 fm$^2$ and presented in Table-IV. The agreement of the present calculations with a wide range of experimental data for the proton radioactivity lifetimes is reasonably good. 
        
\subsection{Alpha radioactivity of SHE}

    The double folded nuclear interaction potential between the daughter nucleus and the emitted particle is given by

\begin{equation}
 V_N(R) = \int \int \rho_1(\vec{r_1}) \rho_2(\vec{r_2}) v_{00}[|\vec{r_2} - \vec{r_1} + \vec{R}|] d^3r_1 d^3r_2 
\label{seqn22}
\end{equation}
\noindent
where $\rho_1$ and $\rho_2$ are the density distribution functions for the two composite nuclear fragments. Since the density dependence of the effective projectile-nucleon interaction was found to be fairly independent of the projectile \cite{Sr83}, as long as the projectile-nucleus interaction was amenable to a single-folding prescription, the density dependent effects on the nucleon-nucleon interaction can be factorized into a target term times a projectile term as 

\begin{equation}
 g(\rho_1, \rho_2) = C (1 - \beta \rho_1^{2/3}) (1 - \beta \rho_2^{2/3}).
\label{seqn23}
\end{equation}   
The parameter $\beta$ can be related to the mean free path in nuclear medium; hence its value should remain the same, 1.5934 fm$^2$, as that obtained from nuclear matter calculations, while the other constant C, which is basically an overall normalization constant, may change. The value of this overall normalization
constant is kept equal to unity, which has been found $\approx$1 from an optimum fit to a large number of $\alpha$ decay lifetimes \cite{Ba03}. This formulation is used successfully in case of $\alpha$ radioactivity of nuclei \cite{Ba03} including superheavies \cite{CSB06,prc07,scb07} and the cluster radioactivity \cite{Ro09}. It can be observed from Eq.(21) that the isovector term becomes zero if anyone (or, both) of the daughter and emitted nuclei involved in the decay process has  N=Z, N and Z being the neutron number and proton number respectively. Therefore in $\alpha$-decay calculations only the isoscalar term contributes. 
            
\noindent 
\begin{table*}[htbp]
\caption{Comparison between experimentally measured and theoretically calculated half-lives of spherical proton emitters. The asterisk symbol (*) in the experimental $Q$ values denotes the isomeric state. The experimental $Q$ values, half lives and $l$ values are from ref. \cite{So02}. The results of the present calculations using the isoscalar and isovector components of DDM3Y folded potentials are compared with the experimental values and with the results of UFM estimates \cite{Bal05}. Experimental errors in $Q$ \cite{So02} values and corresponding errors in calculated half-lives are inside parentheses.}
\begin{tabular}{lllllllll}
\hline
\hline
Parent & $l$ & $Q$ &1$^{st}$ tpt &2$^{nd}$ tpt &3$^{rd}$ tpt &Expt. &This work&UFM       \\ 
$^A Z$& $\hbar$ & MeV &$R_1$[fm]&$R_a$[fm]&$R_b$[fm]&$log_{10}T(s)$&$log_{10}T(s)$& $log_{10}T(s)$ \\ 
\hline
$^{105}Sb$&2&0.491(15)&1.43&6.69&134.30&2.049$^{+0.058}_{-0.067}$&1.90(45)&2.085\\ 
$^{145}Tm$&5&1.753(10)&3.20&6.63&56.27&-5.409$^{+0.109}_{-0.146}$&-5.28(7)&-5.170\\ 
$^{147}Tm$&5&1.071(3)&3.18&6.63&88.65&0.591$^{+0.125}_{-0.175}$&0.83(4)&1.095\\ 
$^{147}Tm^*$&2&1.139(5)&1.44&7.28&78.97&-3.444$^{+0.046}_{-0.051}$&-3.46(6)&-3.199\\ 
$^{150}Lu$&5&1.283(4)&3.21&6.67&78.23&-1.180$^{+0.055}_{-0.064}$&-0.74(4)&-0.859\\ 
$^{150}Lu^*$&2&1.317(15)&1.45&7.33&71.79&-4.523$^{+0.620}_{-0.301}$&-4.46(15)&-4.556\\ 
$^{151}Lu$&5&1.255(3)&3.21&6.69&78.41&-0.896$^{+0.011}_{-0.012}$&-0.82(4)&-0.573\\ 
$^{151}Lu^*$&2&1.332(10)&1.46&7.35&69.63&-4.796$^{+0.026}_{-0.027}$&-4.96(10)&-4.715\\ 
$^{155}Ta$&5&1.791(10)&3.21&6.78&57.83&-4.921$^{+0.125}_{-0.125}$&-4.80(7)&-4.637\\ 
$^{156}Ta$&2&1.028(5)&1.47&7.37&94.18&-0.620$^{+0.082}_{-0.101}$&-0.47(8)&-0.461\\ 
$^{156}Ta^*$&5&1.130(8)&3.21&6.76&90.30&0.949$^{+0.100}_{-0.129}$&1.50(10)&1.446\\ 
$^{157}Ta$&0&0.947(7)&0.00&7.55&98.95&-0.523$^{+0.135}_{-0.198}$&-0.51(12)&-0.126\\ 
$^{160}Re$&2&1.284(6)&1.45&7.43&77.67&-3.046$^{+0.075}_{-0.056}$&-3.08(7)&-3.109\\ 
$^{161}Re$&0&1.214(6)&0.00&7.62&79.33&-3.432$^{+0.045}_{-0.049}$&-3.53(7)&-3.231\\ 
$^{161}Re^*$&5&1.338(7)&3.22&6.84&77.47&-0.488$^{+0.056}_{-0.065}$&-0.75(8)&-0.458\\ 
$^{164}Ir$&5&1.844(9)&3.20&6.91&59.97&-3.959$^{+0.190}_{-0.139}$&-4.08(6)&-4.193\\ 
$^{165}Ir^*$&5&1.733(7)&3.21&6.93&62.35&-3.469$^{+0.082}_{-0.100}$&-3.67(5)&-3.428\\ 
$^{166}Ir$&2&1.168(8)&1.47&7.49&87.51&-0.824$^{+0.166}_{-0.273}$&-1.19(10)&-1.160\\ 
$^{166}Ir^*$&5&1.340(8)&3.22&6.91&80.67&-0.076$^{+0.125}_{-0.176}$&0.06(9)&0.021\\ 
$^{167}Ir$&0&1.086(6)&0.00&7.68&91.08&-0.959$^{+0.024}_{-0.025}$&-1.35(8)&-0.943\\ 
$^{167}Ir^*$&5&1.261(7)&3.22&6.92&83.82&0.875$^{+0.098}_{-0.127}$&0.54(8)&0.890\\ 
$^{171}Au$&0&1.469(17)&0.00&7.74&69.09&-4.770$^{+0.185}_{-0.151}$&-5.10(16)&-4.794\\ 
$^{171}Au^*$&5&1.718(6)&3.21&7.01&64.25&-2.654$^{+0.054}_{-0.060}$&-3.19(5)&-2.917\\ 
$^{177}Tl$&0&1.180(20)&0.00&7.76&88.25&-1.174$^{+0.191}_{-0.349}$&-1.44(26)&-0.993\\ 
$^{177}Tl^*$&5&1.986(10)&3.22&7.10&57.43&-3.347$^{+0.095}_{-0.122}$&-4.64(6)&-4.379\\ 
$^{185}Bi$&0&1.624(16)&0.00&7.91&65.71&-4.229$^{+0.068}_{-0.081}$&-5.53(14)&-5.184\\ \hline
\hline
\label{table4}
\end{tabular} 
\end{table*}

    Comparison between experimental and calculated $\alpha$-decay half-lives for zero angular momenta transfers, using spherical charge distributions for Coulomb interaction and the DDM3Y effective interaction is provided in Table-V. The lower and upper limits of the theoretical half lives corresponding to the upper and lower limits of the experimental $Q_{ex}$ values are also provided. The quantitative agreement with experimental data is reasonable. The results which are underestimated are possibly because the centrifugal barrier required for the spin-parity conservation could not be taken into account due to non availability of the spin-parities of the decay chain nuclei. The term $\hbar^2 c^2 l(l+1) / (2\mu R^2)$ represents the additional centrifugal contribution to the barrier that acts to reduce the tunneling probability if the angular momentum carried by the $\alpha$-particle is non-zero. Hindrance factor which is defined as the ratio of the experimental $T_{1/2}$ to the theoretical $T_{1/2}$ is therefore larger than unity since the decay involving a change in angular momentum can be strongly hindered by the centrifugal barrier. 

\begin{table*}[htbp]
\caption{ Comparison of experimental [Exp] and calculated $\alpha$-decay half-lives using $Q$-values (MeV) from Exp~\cite{oga06,oga04,og04,mo04}, MS~\cite{ms} and M~\cite{Mu01,MU03,Mu03}.}
\begin{center}
\begin{tabular}{ccccccccccccc}
\hline \hline
 Parent &Exp&MS&M&Exp& This Work&This Work&This Work& Ref.  \\ 
\hline
$^A Z$&$Q_{ex}$&$Q^{MS}_{th}$&$Q^{M}_{th}$&$T_{1/2}$&$T_{1/2}[Q_{ex}]$& $T_{1/2}[Q^{MS}_{th}]$&$T_{1/2}[Q^{M}_{th}]$&Exp\\ \hline

$^{294}118$ &$11.81(6)$&12.51&12.11 & $0.89^{+1.07}_{-0.31}$ms&$0.66^{+0.23}_{-0.18}$ms&0.02 ms
&0.15 ms &\cite{oga06}     \\
$^{293}116$ &$10.67(6)$&11.15&11.09& $53^{+62}_{-19} $ms&$206^{+90}_{-61} $ms&12.8 ms&18.3 ms&\cite{og04}     \\  
$^{292}116$  &$10.80(7)$&11.03&11.06&$18^{+16}_{-6} $ms&$39^{+20}_{-13} $ms&10.4 ms&8.65 ms&\cite{og04} \\
$^{291}116$&$10.89(7)$&11.33&10.91&$18^{+22}_{-6} $ms& $60.4^{+30.2}_{-20.1} $ms&5.1 ms&53.9 ms&\cite{oga06} \\  
$^{290}116$&$11.00(8)$&11.34&11.08&$7.1^{+3.2}_{-1.7} $ms&$13.4^{+7.7}_{-5.2} $ms&2.0 ms&8.21 ms&\cite{oga06}  \\ 
$^{288}115$&$10.61(6)$&10.34&10.95&$87^{+105}_{-30} $ms&$410.5^{+179.4}_{-122.7} $ms&2161.6 ms&56.3 ms&\cite{oga04}  \\ 
$^{287}115$&$10.74(9)$&10.48&11.21&$32^{+155}_{-14} $ms&$51.7^{+35.8}_{-22.2} $ms&245.2 ms&3.55 ms&\cite{oga04}  \\ 
$^{289}114$& $9.96(6)$&9.08&10.04&$2.7^{+1.4}_{-0.7} $s&$3.8^{+1.8}_{-1.2} $s&1885 s&2.27 s&\cite{og04}     \\ 
$^{288}114$&$10.09(7)$&9.39&10.32&$0.8^{+0.32}_{-0.18} $s&$0.67^{+0.37}_{-0.27} $s&76.96 s
&0.16 s&\cite{og04}   \\ 
$^{287}114$&$10.16(6)$&9.53&10.56&$0.48^{+0.16}_{-0.09} $s&$1.13^{+0.52}_{-0.35} $s&77.74 s &0.09 s&\cite{oga06}  \\ 
$^{286}114$&$10.33(6)$&9.61&10.86&$0.13^{+0.04}_{-0.02} $s&$0.16^{+0.07}_{-0.05} $s&17.70 s &0.01 s&\cite{oga06} \\ 
$^{284}113$&$10.15(6)$&9.36&10.68&$0.48^{+0.58}_{-0.17} $s&$1.55^{+0.72}_{-0.48} $s&330.19 s &0.06 s&\cite{oga04} \\ 
$^{283}113$&$10.26(9)$&9.56&11.12&$100^{+490}_{-45} $ms&$201.6^{+164.9}_{-84.7} $ms&19845 ms&1.39 ms&\cite{oga04} \\ 
$^{278}113$&$11.90(4)$~$^{a)}$&12.77&---&344 $\mu$s~$^{b)}$&$101^{+27}_{-18}$$\mu$s&1.8 $\mu$s&---&\cite{mo04}  \\
$^{285}112$&$9.29(6)$&8.80&9.49&$34^{+17}_{-9} $s&$75^{+41}_{-26} $s&3046 s&18.6 s&\cite{og04}\\ 
$^{283}112$& $9.67(6)$&9.22&10.16&$3.8^{+1.2}_{-0.7} $s&$5.9^{+2.9}_{-2.0} $s &134.7 s&0.24 s&\cite{oga06}   \\ 
$^{280}111$&$9.87(6)$&10.34&10.77&$3.6^{+4.3}_{-1.3} $s&$1.9^{+0.9}_{-0.6} $s&0.10 s&0.01 s&\cite{oga04}  \\ 
$^{279}111$&$10.52(16)$&11.12&11.08&$170^{+810}_{-80} $ms&$9.6^{+14.8}_{-5.7} $ms&0.34 ms& 0.42 ms&\cite{oga04}  \\ 
$^{274}111$&$11.36(7)$~$^{a)}$&11.07&11.53& 9.26 ms~$^{b)}$&$0.39^{+0.18}_{-0.12} $ms&1.92 ms&0.17 ms&\cite{mo04}  \\ 
$^{279}110$&$9.84(6)$&9.89&10.24&$0.20^{+0.05}_{-0.04} $s&$0.40^{+0.18}_{-0.13} $s&0.29 s&0.03 s&\cite{oga06}\\ 
$^{276}109$&$9.85(6)$&10.11&10.09&$0.72^{+0.87}_{-0.25} $s&$0.45^{+0.23}_{-0.14} $s&0.09 s&0.10 s&\cite{oga04}  \\ 
$^{275}109$&$10.48(9)$&10.26&10.34&$9.7^{+46}_{-4.4} $ms&$2.75^{+1.85}_{-1.09} $ms&10.33 ms&6.36 ms&\cite{oga04}  \\ 
$^{270}109$&$10.23(7)$~$^{a)}$&9.73&10.27&7.16 ms~$^{b)}$&$52.05^{+27.02}_{-17.68} $ms&1235 ms&41.1 ms&\cite{mo04}   \\ 
$^{275}108$&$9.44(6)$&9.58&9.41&$0.19^{+0.22}_{-0.07} $s&$1.09^{+0.61}_{-0.35} $s &0.44 s&1.34 s&\cite{oga06} \\ 
$^{272}107$&$9.15(6)$&9.08&9.08&$9.8^{+11.7}_{-3.5} $s&$10.1^{+5.4}_{-3.4} $s&16.8 s&16.8 s&\cite{oga04}  \\ 
$^{266}107$&$9.26(4)$~$^{a)}$&9.00&8.95& 2.47 s~$^{b)}$&$5.73^{+1.82}_{-1.38} $s&36.01 s&50.8 s&\cite{mo04} \\ 
$^{271}106$&$8.67(8)$&8.59&8.71&$1.9^{+2.4}_{-0.6} $min&$0.86^{+0.71}_{-0.39} $min&1.59 min&0.64 min&\cite{oga06}   \\ 
\hline
\hline
\label{table5}
\end{tabular} 
\end{center}
$a)$ calculated from $\alpha$-decay energies~\cite{mo04}; $b)$ experimental decay times~\cite{mo04}.\\ 
\end{table*}    

    To study the predictive power of the mass formula, $Q$-values are also calculated using the mass formula of Myers and Swiatecki [$Q^{MS}_{th}$] and Muntian et al. [$Q^{M}_{th}$]. Comparison with the theoretical half life values indicates that the $Q$-value predictions of Muntian et al. for the SHE domain are in better agreement with the experimental data than the values obtained from the Myers-Swiatecki mass table.  For example, the theoretical half life of $^{289}114$ obtained using the $Q^{MS}_{th}$  is $\sim 700$ times the experimental one. Where as, calculations in the same framework but with [$Q^{M}_{th}$] as well as experimental $Q_{ex}$-values, agree well over a wide range of experimental data. The theoretical Viola-Seaborg-Sobiczewski (VSS) estimates for $T_{1/2}$ largly overestimates in many cases showing inconsistencies while the present estimate is inconsistent only for few cases where it overestimates but still provides much better estimate than that estimated by VSS systematics. 

\subsection{Cluster radioactivity}

    The decay constant $\lambda$ for cluster radioactivity is a product of cluster preformation probability $P_0$ in the ground state, the tunneling probability through barrier $P$ and the assault frequency $\nu$. The preformation factor may be considered as the overlap of the actual ground state configuration and the configuration representing the cluster coupled to the ground state of the daughter. $P_0$ for normal $\alpha$ emitters is close to unity \cite{bmp93}. Superheavy emitters being loosely bound than highly bound $\alpha$, $P_0$ is expected to be high for $\alpha$ decay and we have seen that the present calculations with $P_0$=1 provide excellent description of $\alpha$ decay for recently discovered superheavy nuclei \cite{CSB06,prc07,scb07}. For weakly bound heavy cluster decay it is expected to be orders of magnitude less than unity. The theoretical half lives of cluster radioactivity for very heavy nuclei are calculated assuming cluster preformation factor to be unity. Hence the preformation factors $P_0$ are calculated \cite{Ro09} as the ratios of the calculated half lives to the experimentally observed half lives which are listed in Table-VI.
       
\noindent
\begin{table}[htbp]
\caption{\label{table2} Half lives assuming preformation factor to be unity and corresponding preformation factors of cluster decay obtained in the present calculation. The asterisk symbol represents theoretical values assuming preformation factor to be unity and using normalisation of 0.7 for the nuclear potentials.} 

\begin{tabular}{cccccccc}
\hline
\hline
 Parent & Cluster &$Q$ (MeV)& $logT_{ex}$ & $logT^*_{th}$ & $logT_{th}$ & $-logP_0$ & $l$   \\ 
\hline
$^{221}$Fr&$^{14}$C&31.317&14.52&13.46&11.71& 2.81 & 3\\ 

$^{221}$Ra&$^{14}$C&32.396&13.39&13.23&11.47& 1.92 & 3\\ 

$^{222}$Ra&$^{14}$C&33.050&11.00&10.38&8.71& 2.29 & 0 \\ 

$^{223}$Ra&$^{14}$C&31.829&15.20&14.34&12.56& 2.64 & 4\\ 

$^{224}$Ra&$^{14}$C&30.540&15.92&15.12&13.34& 2.58 & 0\\ 

$^{225}$Ac&$^{14}$C&30.477&17.34&17.25&15.39& 1.95 & 4\\ 

$^{226}$Ra&$^{14}$C&28.200&21.34&20.22&18.35& 2.99 & 0\\ 

$^{228}$Th&$^{20}$O&44.720&20.72&21.02&18.90& 1.82 & 0\\ 

$^{230}$U&$^{22}$Ne&61.400&19.57&20.81&18.23& 1.34 & 0 \\ 

$^{230}$Th&$^{24}$Ne&57.571&24.64&25.07&22.58& 2.06 & 0\\ 

$^{231}$Pa&$^{24}$Ne&60.417&23.38&22.74&20.32& 3.06 & 1\\ 

$^{232}$U&$^{24}$Ne&62.310&20.40&20.67&18.30& 2.10 & 0\\

$^{233}$U&$^{24}$Ne&60.486&24.82&24.31&21.82& 3.00 & 2\\

$^{234}$U&$^{24}$Ne&58.826&25.25&25.97&23.38& 1.87 & 0\\ 

$^{233}$U&$^{25}$Ne&60.776&24.82&24.35&21.92& 2.90 & 2\\ 

$^{234}$U&$^{26}$Ne&59.466&25.07&26.27&23.84& 1.23 & 0\\ 

$^{234}$U&$^{28}$Mg&74.110&25.74&25.66&22.93& 2.81 & 0\\ 

$^{236}$Pu&$^{28}$Mg&79.670&21.67&21.38&18.80& 2.87 & 0\\ 

$^{238}$Pu&$^{28}$Mg&75.912&25.70&26.25&23.49& 2.21 & 0\\ 

$^{238}$Pu&$^{30}$Mg&76.824&25.28&25.85&23.22& 2.06 & 0\\ 

$^{238}$Pu&$^{32}$Si&91.190&25.30&26.21&23.30& 2.00 & 0\\ 

$^{242}$Cm&$^{34}$Si&96.509&23.15&23.82&21.16& 1.99 & 0\\ \hline
\hline
\label{table6}
\end{tabular} 
\end{table} 

    The basic problem of the present theoretical study is that there is no guarantee that the experimentally observed decays proceed from the ground state of the parent nucleus to that of the daughter nucleus which is assumed in the present calculations. It is a fundamental difficulty associated with the decay of the odd mass parent nuclei or the odd mass emitted clusters. These nuclei may be decaying predominantly to a low-lying excited state of the daughter nucleus. When the exotic cluster is removed from the parent nucleus, the state of the core left over may be quite different from that of the ground state of the daughter nucleus, but rather similar to that of one of its excited states. Hence, decays would go preferentially to this particular excited state, and be strongly suppressed to the ground (and other) states. Present model is not microscopic enough to predict, a priori, which excited state may be most appropriate for this role. However, since the effects due to nuclear structure do not appear explicitly, present calculations can also provide the theoretical half lives for transitions other than ground state to ground state if only the spin-parities and the corresponding experimental $Q$ values are precisely known.

\section{Neutron stars}

\subsection{Modeling neutron Stars}

    If rapidly rotating compact stars were nonaxisymmetric, they would emit gravitational waves in a very short time scale and settle down to axisymmetric configurations. Therefore, we need to solve for rotating and
axisymmetric configurations in the framework of general relativity. For the matter and the spacetime the following assumptions are made. The matter distribution and the spacetime are axisymmetric, the matter and the spacetime are in a stationary state, the matter has no meridional motions, the only motion of the matter is a circular one that is represented by the angular velocity, the angular velocity is constant as seen by a distant observer at rest and the matter can be described as a perfect fluid. The energy-momentum tensor of a perfect fluid $T^{\mu\nu}$ is given by

\vspace{0.0cm}
\begin{equation}
T^{\mu\nu} = (\varepsilon+P)u^\mu u^\nu-g^{\mu\nu}P
\label{seqn24}
\end{equation}
\noindent
where $\varepsilon$, $P$, $u^\mu$ and $g^{\mu\nu}$ are the energy density, pressure, four velocity and the metric tensor, respectively. To study the rotating stars the following metric is used

\vspace{0.0cm}
\begin{eqnarray}
ds^2 = -e^{(\gamma+\rho)} dt^2 + e^{2\alpha} (dr^2+r^2d\theta^2) \nonumber\\
       + e^{(\gamma-\rho)} r^2 \sin^2\theta (d\phi-\omega dt)^2
\label{seqn25}
\end{eqnarray}
\noindent
where the gravitational potentials $\gamma$, $\rho$, $\alpha$ and $\omega$ are functions of polar coordinates $r$ and $\theta$ only. The Einstein's field equations for the three potentials $\gamma$, $\rho$ and $\alpha$ have been solved using the Green's-function technique \cite{Ko89} and the fourth potential $\omega$ has been determined from other potentials. All the physical quantities may then be determined from these potentials. Obviously, at the zero frequency limit corresponding to the static solutions of the Einstein's field equations for spheres of fluid, the present formalism yields the results for the solution of the Tolman-Oppenheimer-Volkoff (TOV) equation \cite{TOV39}. We use the `rns' code \cite{St95} for calculating the compact star properties which requires EoS in the form of energy density versus pressure along with corresponding enthalpy and baryon number density and since we are using various EoS for different regions, these are smoothly joined.  

    The different regions of a compact star are governed by different EoS. These can be broadly divided into two regions: a crust that accounts for about 5$\%$ of mass and about 10$\%$ of the radius of a star and the core is responsible for the rest of the mass and radius of a star. The outer layers are a solid crust $\sim$ 1 km thick, consisting, except in the outer few meters, of a lattice of bare nuclei immersed in a degenerate electron gas. As one goes deeper into the crust, the nuclear species become, because of the rising electron Fermi energy, progressively more neutron rich, beginning (ideally) as $^{56}$Fe through $^{118}$Kr at mass density $\approx$4.3$\times$10$^{11}$ g cm$^{-3}$. At this density, the `neutron drip' point, the nuclei have become so neutron rich that with increasing density the continuum neutron states begin to be filled, and the lattice of neutron-rich nuclei becomes permeated by a sea of neutrons.
    
    The EoS that cover the crustal region of a compact star are Feynman-Metropolis-Teller (FMT) \cite{FMT49}, Baym-Pethick-Sutherland (BPS) \cite{BPS71} and Baym-Bethe-Pethick (BBP) \cite{BBP71}. The most energetically favorable nucleus at low densities is $^{56}$Fe, the endpoint of thermonuclear burning. The FMT is based on Fermi-Thomas model to derive the EoS of matter at high pressures and covers the outermost crust which is essentially made up of iron and a fraction of the electrons bound to the nuclei. The major difficulty in deriving the equation of state is the calculation of the electronic energy. At subnuclear densities, from about 10$^4$ g cm$^{-3}$ up to the neutron drip density 4.3$\times$10$^{11}$ g cm$^{-3}$ the EoS of BPS is applicable which includes the effects of the lattice Coulomb energy on the equilibrium nuclide. The domain from neutron drip density to about nuclear density 2.5$\times$10$^{14}$ g cm$^{-3}$, is composed of nuclei, electrons and free neutrons where EoS of BBP is applicable which is based on a compressible liquid drop model of nuclei with conditions that nuclei must be stable against $\beta$-decay and free neutron gas must be in equilibrium with neutrons in nuclei. 
    
\begin{figure}[t]
\vspace{0.0cm}
\centerline{\epsfig{file=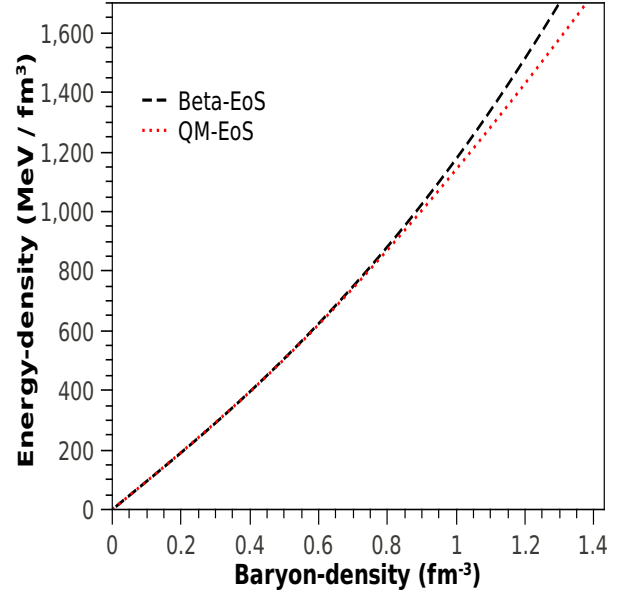,height=8.0cm,width=8.0cm}}
\caption
{ The EoS of the $\beta$-equilibrated charge neutral neutron star matter and the quark matter EoS.}
\label{fig7}
\vspace {1.7cm}
\end{figure}

\begin{figure}[t]
\vspace{0.0cm}
\centerline{\epsfig{file=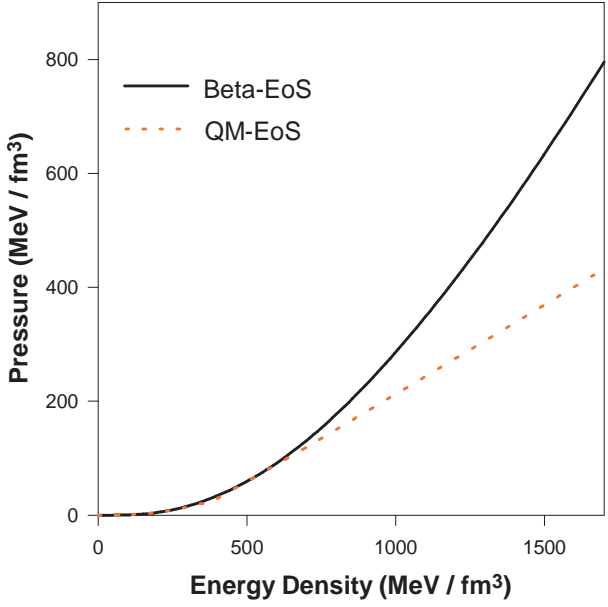,height=8.0cm,width=8.0cm}}
\caption
{Pressure versus energy density plots of $\beta$-equilibrated charge neutral neutron star matter and quark matter.}
\label{fig8}
\vspace {0.0cm}
\end{figure}
    
\begin{figure}[t]
\vspace{0.0cm}
\centerline{\epsfig{file=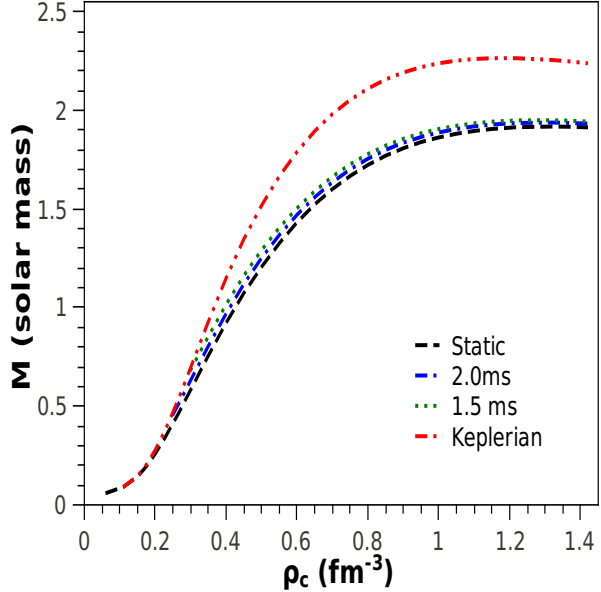,height=8.0cm,width=8.0cm}}
\caption
{ Variation of mass with central density for static and rotating neutron stars with pure nuclear matter inside.}
\label{fig9}
\vspace {0.0cm}
\end{figure}

\begin{figure}[t]
\vspace{0.0cm}
\centerline{\epsfig{file=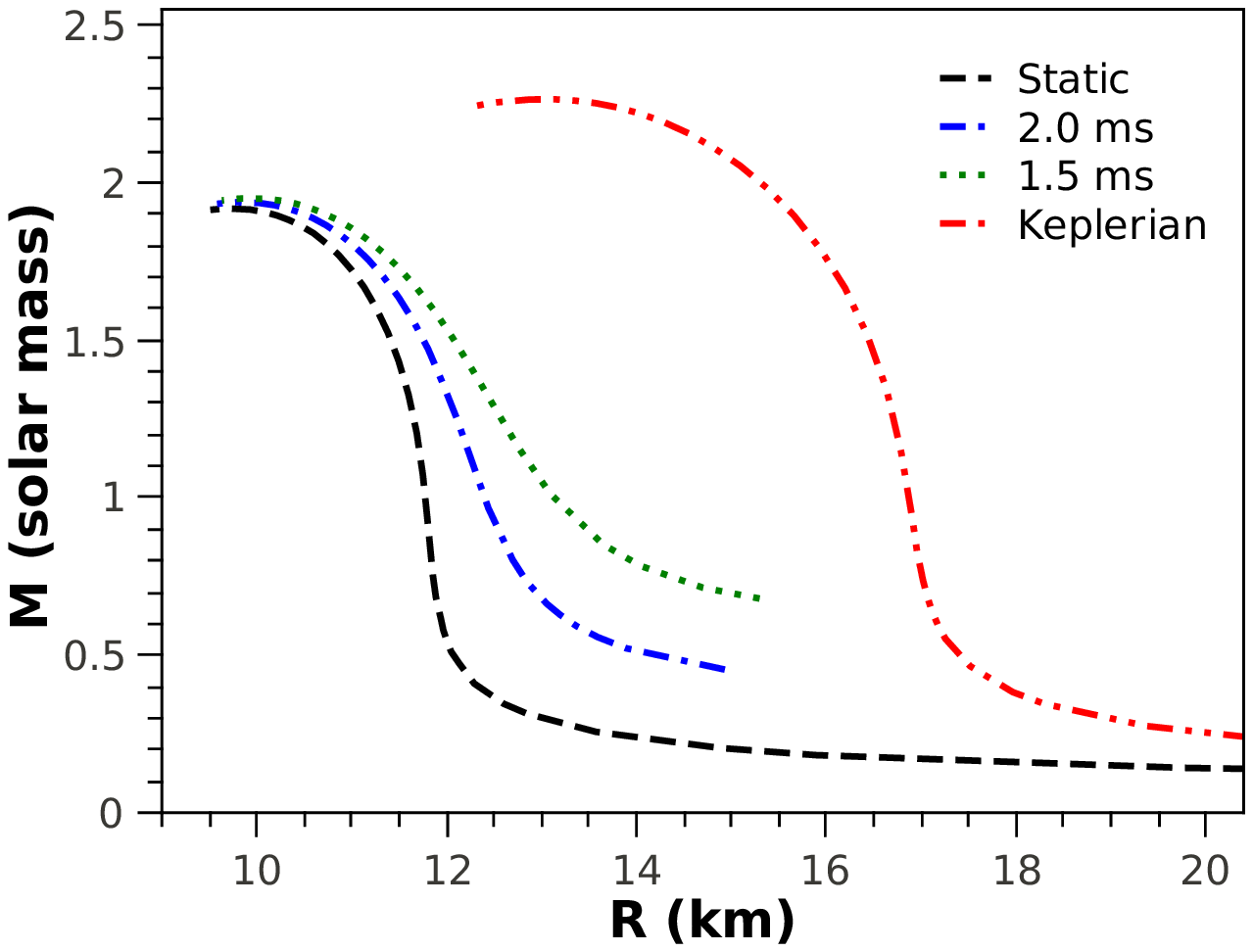,height=8.0cm,width=8.0cm}}
\caption
{ Mass-radius(equatorial) relationship for static and rotating neutron stars with pure nuclear matter inside.}
\label{fig10}
\vspace{0.0cm}
\end{figure}

\subsection{$\beta$-equilibrated neutron star matter and quark matter EoS}

    The nuclear matter EoS,  as described earlier, is calculated \cite{BCS08} using the isoscalar and the isovector components of M3Y interaction along with density dependence which is completely determined from the nuclear matter calculations. This EoS evaluated at the isospin asymmetry $X$ determined from the $\beta$-equilibrium proton fraction $x_\beta$ [$=\frac{\rho_p}{\rho}$], obtained by solving 

\begin{equation}
 \hbar c (3 \pi^2\rho x_\beta)^{1/3}= 4E_{sym}(\rho) (1 - 2 x_\beta),
\label{seqn26}
\end{equation}
\noindent
provides EoS for the $\beta$-equilibrated NS matter (Fig.-7) where $E_{sym}(\rho)$ is the nuclear symmetry energy.
  
    For cold and dense quark (QCD) matter, the perturbative EoS \cite{Ku10} with two massless and one massive quark flavors and a running coupling constant, is used. The constant $B$ is treated as a free parameter, which allows to take into account non-perturbative effects not captured by the weak coupling expansion. In fact, using the free quark number density, one recovers the expression for the pressure in the original MIT bag model \cite{Ch74}, with $B$ taking the role of the bag constant. Due to physics criteria (e.g. requiring the energy density to be positive), the possible values for $B$ are, however, typically rather restricted, allowing to make quantitative statements that are not possible in the original MIT bag model.  

    The energy density of the quark matter is lower than that of the present EoS for the $\beta$-equilibrated charge neutral NS matter at densities higher than 0.405 fm$^{-3}$ for bag constant $B^{\frac{1}{4}}$=110 MeV \cite{Ku10} implying presence of quark core. The energy densities of the present EoS for the $\beta$-equilibrated charge neutral NS matter and the quark matter EoS for bag constant $B^{\frac{1}{4}}$=110 MeV are shown in Fig.-7 as functions of baryonic densities. For lower values of bag constant such as $B^{\frac{1}{4}}$=89 MeV, energy density for our EoS is lower and makes a cross over with the quark matter EoS at very high density $\sim$1.2 fm$^{-3}$ causing too little quark core (predicting similar results as NS with pure nuclear matter inside) and therefore we choose $B^{\frac{1}{4}}$=110 MeV for representative calculations. This means that the value of $B^{\frac{1}{4}}$=110 MeV is arbitrarily chosen to allow for a phase transition at densities that are reachable in the core of neutron stars. Since the energy density $\varepsilon=\rho(\epsilon+m c^2)$, $\frac{d\varepsilon}{d\rho}=\epsilon+m c^2+\rho \frac{d\epsilon}{d\rho}$ implying $\varepsilon=\rho\frac{d\varepsilon}{d\rho}-P$. Thus the negative intercept of the tangent (having slope $\frac{d\varepsilon}{d\rho}$) drawn at a point to the energy density versus density plot represents pressure $P=\rho^2\frac{d\epsilon}{d\rho}$ at that point. The common tangent is drawn for the energy density versus density plots where pressure is the negative intercept of the tangent to energy density versus density plot. However, as obvious from Fig.-7, the phase co-existence region is negligibly small which is represented by part of the common tangent between the points of contact on the two plots \cite{Pe93} implying constant pressure throughout the phase transition. This is, thus, equivalent to Maxwell's construction. In Fig.-8, plots of pressure as a function of energy density are shown for $\beta$-equilibrated charge neutral neutron star matter and quark matter (corresponding to bag constant $B^{\frac{1}{4}}$=110 MeV).

\begin{figure}[t]
\vspace{0.0cm}
\centerline{\epsfig{file=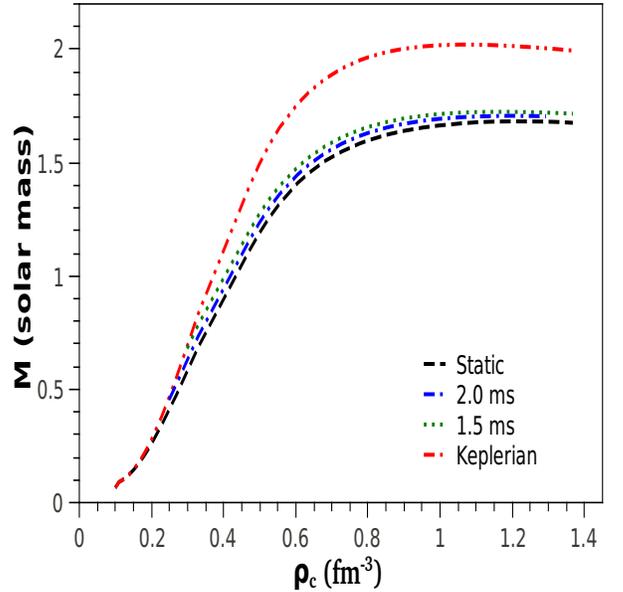,height=8.0cm,width=8.0cm}}
\caption
{ Variation of mass with central density for static and rotating neutron stars with nuclear and quark matter inside.}
\label{fig11}
\vspace {0.0cm}
\end{figure}

\begin{figure}[t]
\vspace{0.0cm}
\centerline{\epsfig{file=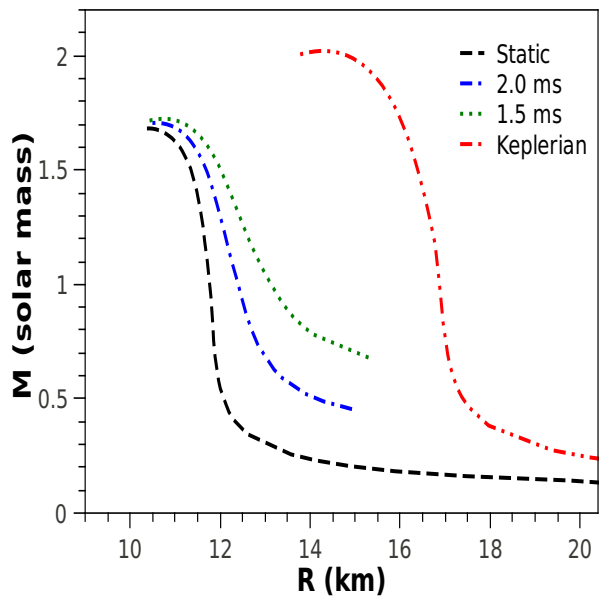,height=8.0cm,width=8.0cm}}
\caption
{ Mass-radius(equatorial) relationship for static and rotating neutron stars with nuclear and quark matter inside.}
\label{fig12}
\vspace{0.0cm}
\end{figure}

\subsection{Calculations and results: masses and radii of neutron and hybrid Stars}

    The rotating compact star calculations are performed using the crustal EoS, FMT + BPS + BBP upto number density of 0.0458 fm$^{-3}$ and $\beta$-equilibrated NS matter beyond. It is worthwhile to mention here that a star may not rotate as fast as Keplerian frequency due to r-mode instability. There have been suggestions that the r-mode instability may limit the time period to 1.5 ms \cite{St06}. However, a pulsar rotating faster (e.g., PSR J17482446ad) than this limit has already been observed \cite{He06}. The variation of mass with central density for static and rotating neutron stars at Keplerian limit and also maximum frequencies limited by the r-mode instability with pure nuclear matter inside is shown in Fig.-9. In Fig.-10, the mass-radius relationship for static and rotating neutron stars at Keplerian limit and also at maximum frequencies limited by the r-mode instability with pure nuclear matter inside is shown. Fig.-10 depicts that NSs with pure nuclear matter inside, the maximum mass for the static case is 1.92 M$_\odot$ with radius $\sim$9.7 km and for the star rotating with Kepler's frequency it is 2.27 M$_\odot$ with equatorial radius $\sim$13.1 km \cite{Ch10}. However, for stars rotating with maximum frequency limited by the r-mode instability, the maximum mass turns out to be 1.95 (1.94) M$_\odot$ corresponding to rotational period of 1.5 (2.0) ms with radius about 9.9 (9.8) kilometers.

    The variation of mass with central density for static and rotating neutron stars at Keplerian limit and also maximum frequencies limited by the r-mode instability with nuclear and quark matter inside is shown in Fig.-11. In Fig.-12, the mass-radius relationship for static and rotating neutron stars at Keplerian limit and also at maximum frequencies limited by the r-mode instability with nuclear and quark matter inside is shown. Fig.-12 depicts that when quark core is considered, the maximum mass for the static case is 1.68 M$_\odot$ with radius $\sim$10.4 km and for the star rotating with Kepler's frequency it is 2.02 M$_\odot$ with equatorial radius $\sim$14.3 km. In a similar study, it was concluded that compact stars with a quark matter core and an hadronic outer layer, can be as massive as 2.0 M$_\odot$ but stay below the pure quark stars and pure neutron stars \cite{We11}. However, they have used two different relativistic mean-field parameter sets TM1 and NL3 \cite{We11} to explore the influence of the hadronic part of the EoS whose high density behaviour do not satisfy the criteria extracted from the experimental flow data \cite{Da02}. For our case stars rotating with maximum frequency limited by the r-mode instability, the maximum mass turns out to be 1.72 (1.71) M$_\odot$ corresponding to rotational period of 1.5 (2.0) ms with radius about 10.7 (10.6) kilometers \cite{APP12}. 

\section{Summary and conclusion}

    In summary, we show that theoretical description of nuclear matter based on mean field calculation using density dependent M3Y effective NN interaction yields a value of nuclear incompressibility which is highly in agreement with that extracted from experiment and gives a value of NSE that is consistent with the empirical value extracted by fitting the droplet model to the measured atomic mass excesses and with other modern theoretical descriptions of nuclear matter. The present NSE is `soft' because it increases initially with nucleonic density up to about two times the normal nuclear density and then decreases monotonically at higher densities and is consistent with the recent evidence for a soft NSE at suprasaturation densities \cite{Zh09}. The slope $L$ and the isospin dependent part $K_\tau$ of the isobaric incompressibility are consistent with the constraints recently extracted from analyses of experimental data. Of all other models, DD-ME2 provides comparatively better estimates and of all RMF models provides best mass predictions with r.m.s. deviation of 0.9 MeV. Interestingly, our calculations provide so far the best theoretical estimates for the isospin dependent properties of asymmetric nuclear matter.

    The maximum of the $\beta$-equilibrium proton fraction $x_\beta\approx0.044 $ calculated using the present NSE occurs at $\rho\approx1.35\rho_0$ and goes to zero at $\rho \approx 4.5\rho_0$ and therefore forbids the direct URCA process since the equilibrium proton fraction is always less than 1/9 \cite{La91}. This feature is consistent with the fact that there are no strong indications \cite{AWS06,Ca06} that fast cooling occurs. It was also concluded theoretically that an acceptable EoS of asymmetric nuclear matter shall not allow the direct URCA process to occur in neutron stars with masses below 1.5 solar masses \cite{Kl06}. Even recent experimental observations that suggest high heat conductivity and enhanced core cooling process indicating the enhanced level of neutrino emission, were not attributed to the direct URCA process but were proposed to be due breaking and formation of neutron Cooper pairs \cite{Cr10,Da11,Dm11,Sh11}.

    The recently measured data on the breathing mode of Sn isotopes seem to favour a constraint $K_\tau= -550\pm100$ MeV for the asymmetry term in the nuclear incompressibility \cite{Li07,Ga07}. First and foremost, $K_\tau$ should not be inferred from an extrapolation to the $A \rightarrow \infty$ limit from laboratory experiments on finite nuclei. Rather, one should continue to follow the procedure advocated by Blaizot \cite{Bl80,Bl95} and demand that the values of both $K_\infty$ and $K_\tau$ be those predicted by a consistent theoretical model that successfully reproduces the experimental giant monopole resonance (GMR) energies of a variety of nuclei. We reiterate that in the present contribution, both $K_\infty$ and $K_\tau$ refer to bulk properties of the infinite system. Nevertheless, considering the fact that the extracted value of $K_\tau=-550\pm100$ MeV \cite{Li07} is from GMR of nuclei as light as Sn isotopes, the present value $-408.97\pm3.01$ MeV is in reasonably close agreement whereas it is in excellent agreement with $K_\tau=-389\pm12$ MeV (NL3),$-345\pm12$ MeV (SVI2),$-395\pm13$ MeV (SIGO-c) \cite{Sh09} when extracted reproducing GMR energies of nuclei such as $^{208}$Pb, Sn isotopes and $^{90}$Zr among others.

    The energy density of the present EoS for $\beta$-equilibrated charge neutral NS matter using DDM3Y effective NN interaction turns out to be higher than that of quark matter at densities above 0.405 fm$^{-3}$ implying possibility of quark core. We have applied our nucleonic EoS with FMT + BPS + BBP crustal EoSs, to solve the Einstein's field equations to determine the mass-radius relationship of neutron stars with and without quark cores. The result for NS without quark core is in excellent agreement with recent astrophysical observations. The neutron star matter can further undergo deconfinement transition to quark matter, thereby reducing compact star masses considerably. Although pure NSs rotating with maximum frequency limited by the r-mode instability have masses up to $\sim$2 M$_\odot$, but hybrid compact stars rotating with maximum frequency limited by the r-mode instability (or with 3.1 ms as observed for pulsar J1614-2230), the maximum mass $\sim$1.7 M$_\odot$ turns out to be lower than the observed mass of 1.97$\pm$0.04 M$_\odot$ and thus rules out quark cores for such massive pulsars but not for pulsars with masses $\sim$1.7 M$_\odot$ or less. Obviously, in order not to conflict with the mass measurement \cite{De10}, either there must be some mechanism to prevent nuclear matter to deconfine into quark matter or the quark EoS should be made stiffer by several possible realistic improvements \cite{Ku10}. The nucleon-nucleon effective interaction used in the present work, which is found to provide a unified description of elastic and inelastic scattering, various radioactivities and nuclear matter properties, also provides an excellent description of the $\beta$-equilibrated NS matter which is stiff enough at high densities to reconcile with the recent observations of the massive compact stars $\sim$2 M$_\odot$ while the corresponding symmetry energy is supersoft \cite{CBS09} as preferred by the FOPI/GSI experimental data.

\end{document}